\documentclass[preprint2]{aastex}

\usepackage{natbib}
\usepackage{epsfig}
\usepackage{amssymb,amsmath}
\usepackage{color}
\usepackage{xspace}
\usepackage{multirow}
\usepackage{bigstrut}
\usepackage{longtable}

\def\ifm#1{\relax\ifmmode#1\else$\mathsurround=0pt #1$\fi}

\def\ltsima{$\; \buildrel < \over \sim \;$}
\def\lsim{\lower.5ex\hbox{\ltsima}}
\def\gtsima{$\; \buildrel > \over \sim \;$}
\def\gsim{\lower.5ex\hbox{\gtsima}}

\def\ie{{\it i.e.}\,}

\newcommand{\kms}{\, {\rm km\, s}^{-1}}

\newcommand{\kpc}{\, {\rm kpc}}

\newcommand{\mypm}[2]{^{+#1}_{-#2}}

\newcommand{\rsrc}{R_{\rm s}}
\newcommand{\logt}{\log_{\rm 10}}

\def\zd{z_{\rm d}}
\def\zs{z_{\rm s}}

\def\Reff{R_{\rm eff}}
\def\Reff2{R_{\rm eff}/2}
\def\Reff{R_{\rm eff}}
\def\REin{R_{\rm Ein}}

\def\fdmsre2{f^{\rm Salp}_{{\rm DM,}\Reff2}}
\def\fdmcre2{f^{\rm Chab}_{{\rm DM,}\Reff2}}

\def\hst{{\it HST}{ }}

\def\sigmae2{\sigma_{\rm e2}}

\def\RF{{\sc RingFinder}\xspace}

\def\ie{{\it i.e.}\xspace}

\def\Sref#1{Section~\ref{#1}\xspace}
\def\Fref#1{Figure~\ref{#1}\xspace}
\def\Tref#1{Table~\ref{#1}\xspace}
\def\Eref#1{Equation~\ref{#1}\xspace}

\def\gmi{(g-i)}

\def\imax{22}

\def\Nf{330}
\def\NfX{71}

\def\NETGs{640,000\xspace}
\def\NRF{2500\xspace}
\def\XRF{0.4\%\xspace} 
\def\Ncands{\Nf\xspace}

\def\Nlens{33\xspace}
\def\Nlensn{16\xspace} 
\def\SArea{150\xspace}

\def\Nsim{96,\,000\xspace}

\def\iap{1}
\def\kipac{2}
\def\oxford{3}
\def\ucsb{4}

\shorttitle{Automated Lens Detection}
\shortauthors{Gavazzi et~al.}

\begin{document}

\title{\RF: automated detection of galaxy-scale gravitational lenses \\
       in ground-based multi-filter imaging data}

\author{Rapha\"el~Gavazzi\altaffilmark{\iap}}
\author{Philip~J.~Marshall\altaffilmark{\kipac,\oxford}}
\author{Tommaso~Treu\altaffilmark{\ucsb}}
\author{Alessandro~Sonnenfeld\altaffilmark{\ucsb}}

\altaffiltext{\iap}{Institut d'Astrophysique de Paris, UMR7095 CNRS -- Universit\'e Pierre et Marie Curie, 98bis bd Arago, 75014 Paris, France} 
\altaffiltext{\kipac}{Kavli Institute for Particle Astrophysics and Cosmology, P.O.\ Box 20450, MS29, Stanford, CA 94309, USA}
\altaffiltext{\oxford}{Department of Physics, Oxford University, Denys Wilkinson Building, Keble Road, Oxford OX1 3RH, UK}
\altaffiltext{\ucsb}{Physics department, University of California, Santa Barbara, CA 93601, USA}


\begin{abstract}  
We present \RF, a tool for finding galaxy-scale strong gravitational lenses in multi-band imaging data.
By construction, the method is sensitive to configurations involving a massive foreground early-type galaxy
and a faint, background, blue source. \RF~detects the presence of blue residuals embedded in an otherwise smooth red light
distribution by difference imaging in two bands. The method is automated for efficient application to current and future surveys,
having originally been designed for the 150-deg2 Canada France Hawaii Telescope Legacy Survey (CFHTLS).
We describe each of the steps of \RF. We then carry out extensive simulations to assess completeness and purity.
For sources with magnification $\mu>$4, \RF~reaches 42\% (resp. 25\%) completeness and 29\% (resp. 86\%) purity before
(resp. after) visual inspection. The completeness of \RF~is substantially improved in the particular range of Einstein radii
$0\farcs 8 \le \REin \le 2\farcs0$ and lensed images brighter than $g=22.5$, where it can be as high as $\sim$70\%.
\RF~does not introduce any significant bias in the source or deflector population.
We conclude by presenting the final catalog of \RF~CFHTLS galaxy-scale strong lens candidates. Additional information
obtained with Hubble Space Telescope and Keck Adaptive Optics high resolution imaging, and with Keck and Very Large Telescope spectroscopy,
is used to assess the validity of our classification, and measure the redshift of the foreground and the background objects.
From an initial sample of 640,000 early type galaxies, \RF~returns 2500 candidates, which we further reduce by visual
inspection to 330 candidates. We confirm 33 new gravitational lenses from the main sample of candidates,
plus an additional 16 systems taken from earlier versions of \RF.
First applications are presented in the SL2S galaxy-scale Lens Sample paper series.
\end{abstract}



\section{Introduction}\label{sec:intro}

Since the discovery of the first multiple quasar produced by  strong
gravitational lensing by a foreground massive galaxy \citep{walsh79},
and the discovery of the first giant arcs found at the centers of
galaxy clusters \citep{soucail87,lynds86}, much progress has been made
in exploiting the unique capabilities of strong gravitational lensing
as a probe of the mass content of distant massive objects, independent
of the nature of their constituents or their dynamical state.  With
the advent of deep, wide-field optical imaging surveys we have now
entered a new era that enables the use of sizable samples of strong
lensing events as precision diagnostics of the physical properties of
the distant Universe.

Gravitational lensing, by itself and in combination with other probes,
can be used to great effect to measure the mass profiles of early-type
galaxies, both in the nearby universe and at cosmological distances
\citep[e.g.\ ][]{T+K02a,T+K02b,RKK03,T+K04,R+K05,Koo++06,J+K07,Gav++07,Tre10,Aug++10,Lag++10,Son++12,Bol++12,Dye++13,ORF13}.
Until recently, however, this approach was severely limited by the
small size of the samples of known strong gravitational lenses. This
has motivated a number of dedicated searches which have increased the
sample of known strong gravitational lens systems by more than an
order of magnitude in the past decade. 

Different search strategies have been adopted depending on the
properties of the parent survey. A fundamental distinction is that
between source-oriented and deflector-oriented searches
\citep[e.g.][]{SEF92}. The choice depends on the relative abundance of
the population of foreground deflectors and that of background
sources, and has strong implications for the potential applications of
the resulting lens catalog.

Historically, source-oriented surveys were considered first, as they
generally require the analysis of a relatively small input catalog of
magnified sources that turn out to be much brighter than the
foreground deflector at a carefully chosen wavelength.  The sources of
choice were typically bright quasars or radiosources, that would
outshine the light from the deflector. This approach is
well-illustrated in the optical by the Sloan Quasar Lens Survey
\citep[SQLS][]{Ina++03,Ina++12}, the near IR with MUSCLES
\citep{Jac++12}, and in the radio by the Cosmic Lens All Sky Survey
(CLASS) \citep{Mye++03,Bro++03}. This last survey led to the early
discovery of 22 gravitationally lensed quasars, many of which have
been followed up with \hst. The latest release of the SQLS lens
catalog has reported the discovery of 49 new quasar lenses.
The scarcity of the population of background sources 
\citep[$\sim0.1$ per deg$^2$,][]{O+M10} requires extremely wide field
surveys in order to gather sizable lens samples.

Recently, wide field imaging surveys at millimeter wavelengths with
the South Pole Telescope \citep[SPT,][]{Hez++13}, and sub-millimeter
wavelengths with the {\it Herschel} satellite like H-ATLAS
\citep{Neg++10,Gon++12,Bus++13} and HerMES \citep{Con++11,Gav++11,War++13},
have made it possible to target the population of distant sub-mm galaxies
in the redshift range $1-4$ and find large number of lenses, a result predicted by
\citet{Bla96}.

These surveys lead to a spatial density of strong lenses ranging
between 0.1 and 0.2 deg$^{-2}$ \citep{Neg++10,War++13,Vie++13}. The
recent availability of high-resolution spectro-imaging with the
Atacama Large Millimeter Array (ALMA) of gravitational lenses found at
those wavelengths makes this technique a very promising avenue for the
coming decade \citep{Hez++13}. We expect that a similar density of
lensed quasars will be reached by optical surveys as well, including
the Dark Energy Survey, the HSC Survey and in the next decade LSST and
Euclid \citep{O+M10}.

Indeed, at optical wavelengths, where unobscured  star-forming
galaxies are clearly visible, the ever-increasing deep wide field
imaging and spectroscopic surveys are providing a population of
distant background sources that has reached several hundreds of
thousands per square degree.  Since these are generally much fainter
than the foreground massive early type galaxy deflectors, an effective
strategy is to focus on these less numerous foreground galaxies and
look for signatures of a gravitationally-lensed background object.
As most of these background sources are spatially resolved, they take
the typical shape of a complete, or partial, arc-like, Einstein ring.
The challenge of this approach generally resides in the limited
spatial resolution of wide field surveys, and the somewhat similar
wavelengths at which both the lens and the source shine, which makes
it difficult to disentangle the source and deflector light.

A particularly successful way to mitigate this problem has been to
take advantage of large spectroscopic surveys, such as the Sloan
Digital Sky Survey (SDSS), which took spectra of several hundreds of
thousands of bright galaxies. By looking for composite spectra
consisting of two objects at different redshifts within the solid
angle covered by the spectroscopic fiber, it has been possible to
build a large sample of galaxy-galaxy lens systems. These consist
typically of a low redshift foreground massive early-type galaxy and a
background star-forming galaxy at higher redshift. The Sloan Lens ACS
Survey (SLACS) conducted \hst follow-up observations of such
spectroscopic candidates, and discovered about 100 gravitational
lenses in the redshift range $0.1\le \zd\le 0.4$
\citep{Bol++04,Bol++06,Aug++10}. Rarer configurations, involving a
foreground late-type galaxy, or a background early-type galaxy, were
also searched for \citep{Tre++11,Bre++12,Aug++11}. Recently, this
technique has been extended to the SDSS-III survey and has been used
to find more lenses at $z\lesssim 0.6$ \citep{Bro++12,Bol++12}. The
main advantage of this spectroscopic approach is that important
quantities, such as the deflector and source redshifts along with the
velocity dispersion of the stars in the deflector, are obtained from
the parent survey itself. The availability of this data allows for
many scientific applications including, combined lensing+dynamical
studies of these systems
\citep{T+K04,Koo++06,Koo++09,Aug++10,Son++12}, without the need
for targeted spectroscopic follow-up.

Even though the approaches described above have been very successful,
there is strong motivation to develop techniques to identify
galaxy-galaxy lenses purely in imaging data. This is challenging, but
it can potentially yield a larger number of objects than any other
technique: \citet{MBS05} forecast more than 10 such systems per square
degree at \hst-like depth and resolution. Similar numbers are expected
for Euclid/LSST. We therefore expect an all-sky survey should find
more than $10^5$ such systems. Finding a similar sample of systems
from an all-sky spectroscopic catalog would require of order $10^8$
spectra, two orders of magnitude more than have been taken to date. 

Because of the difficulty of identifying galaxy-galaxy lenses in
optical images, much effort has been devoted to the analysis of \hst
data in order to exploit its resolution. Searches based on both visual
and automated inspection have been conducted
\citep[e.g.][]{RGO99,Mou++07,Fau++08,Jac08,Mar++09,Paw++12,NMT09},
yielding several tens of candidates over the few square degrees of
available data.

Still, ground-based imaging is a potentially promising avenue, with
its relatively low angular resolution compensated by the ready
availability of hundreds or thousands of square degrees in multiple
bands. In the SDSS, with typically $1\farcs5$ seeing and limiting
magnitude $r\sim 21.5$, only wide separation systems produced by very
massive galaxies, groups or clusters of galaxies have been able to be
uncovered \citep{Bel++09}. The success rate increases significantly 
with angular resolution, and therefore sub-arcsecond image quality is
desirable. Thus, good image quality and wide area coverage, such as
that provided by the Canada France Hawaii Legacy Survey (CFHTLS), are
potentially more promising sources of lenses -- especially those with
deflector redshifts $\sim0.5$ and above, where current samples are
scant \citep{Tre++10}. This argument motivated the Strong Lensing
Legacy Survey (SL2S), which comprised a search for group and cluster
scale (Einstein Radius $\REin\gtrsim 3\arcsec$) lenses
\citep{Cab++07,Mor++12}, and the
present work, the SL2S galaxy-scale lens search. The ultimate goal of
our work was to use the newly found lenses to study the formation and
evolution of massive galaxies; our results can be found in 
\citet{Ruf++11,Gav++12,PaperIII,PaperIV}.

In this paper we present \RF, the semi-automated procedure for finding
galaxy-scale strong lenses in the multi-band imaging data that we have
applied to the CFHTLS. By focusing on the most frequent lens-source
configuration, that of a foreground red massive early-type galaxy and
a faint blue background source at higher redshift, we implement a
technique that subtracts off the foreground light and analyses the
blue residuals by requiring that they are broadly consistent with a
strong lensing event. The CFHTLS data are described in
\Sref{sect:data} while \Sref{sect:method} describes the
\RF~algorithm. A list of \Nf~lens candidates, plus \NfX~additional
candidates from preliminary versions of \RF~and additional datasets
are also presented in this section.

In Section~\ref{sect:simus} we describe the extensive and realistic
simulations of plausible galaxy-scale lens systems which we use to
assess the performance of the method, both in terms of completeness
and purity. We explore the dependence of those quantities on important
parameters like the Einstein radius, source magnitude, and lens and
source redshifts in order to characterize the selection function of
the \RF sample.

Then,  in Section~\ref{sect:followup} we summarize the results of our
multi-year follow-up campaign with high resolution imagers and
spectrographs, present the sample of confirmed lenses, and discuss
false positives and contaminants.  We summarize our main results and
present our conclusions in Section~\ref{sect:concl}.

Throughout this paper, all magnitudes referred to are calculated in
the AB system, and we assume the concordance $\Lambda$CDM cosmological
background with $\Omega_{\rm m}=0.3$, $\Omega_\Lambda=0.7$ and
$H_0=70$ km s$^{-1}$ Mpc$^{-1}$.


\section{The CFHT Legacy Survey}\label{sect:data} 

The Canada-France-Hawaii Telescope Legacy
Survey\footnote{\url{http://www.cfht.hawaii.edu/Science/CFHLS}}
(CFHTLS) is a major photometric survey of more than 450 nights over 5
years (started on June 1st, 2003) using the MegaCam wide field imager
which covers $\sim$1 square degree on the sky, with a pixel size of
0\farcs186. The CFHTLS has two components aimed at extragalactic
studies: a Deep component consisting of 4 pencil-beam fields of
1\,deg$^2$ and a Wide component consisting of 4 mosaics covering
\SArea\,deg$^2$ in total. Both surveys are imaged through 5 broadband filters.  The data are
pre-reduced at CFHT with the Elixir pipeline\footnote{\tt
http://www.cfht.hawaii.edu/Instruments/Elixir/} which removes the
instrumental artifacts in individual exposures. The CFHTLS images are
then astrometrically calibrated, photometrically inter-calibrated,
resampled and stacked by the Terapix group at the Institut
d'Astrophysique de Paris (IAP) and finally archived at the Canadian
Astronomy Data Centre (CADC). Terapix also provides weight map images,
quality assessments meta-data for each stack as well as mask files
that mask saturated stars and defects of each image. The production of
photometric catalogs was made at Terapix using {\tt SExtractor}
\citep{B+A96}. In this paper we use the sixth data release (T0006) described in detail by \citep{Gor++09}\footnote{see
also,
\url{http://terapix.iap.fr/rubrique.php?id\_rubrique=259},
\url{http://terapix.iap.fr/cplt/T0006-doc.pdf}}


The Wide survey is a
single epoch imaging survey, covering some \SArea deg$^2$ in 4 patches
of the sky. It reaches a typical depth of $ u^*\simeq 25.35$, $
g\simeq 25.47$, $r\simeq24.83$, $i\simeq 24.48$ and $ z\simeq 23.60$
(AB mag of 80\% completeness limit for point sources) with typical
FWHM point spread functions of $0\farcs85$, $0\farcs79$, $0\farcs71$,
$0\farcs64$ and $0\farcs68$, respectively. Because of the greater
solid angle, the Wide component is our main provider of lens
candidates and, unless otherwise stated, the analysis below refers to
the application of \RF~to the Wide survey.

Regions around the halo of bright saturated stars, near CCD defects or
near the edge of the fields have lower quality photometry and are
discarded from the analysis. Overall, we reject $\sim 21\%$ of the
CFHTLS Wide survey area, reducing the total area analyzed in this work
to 135.2 deg$^2$.


\section{The \RF~pipeline}
\label{sect:method} 


In this section we present the methodology that we adopted to uncover
gravitational lenses in multi-filter imaging data. It is a
lens-oriented strategy in the sense that we first select bright
early-type galaxies (ETGs) as they presumably are the most massive
objects and thus the most efficient gravitational lenses. Then, the
bulk of the deflector light is removed by subtracting a scaled version
of the $i$-band image from the $g$-band image. Finally, we look for
blue features consistent with being gravitationally lensed objects in
the residuals.


\subsection{Population of foreground Early-Type Galaxies}
\label{sect:method:selection}

The starting point of our procedure is a photometric catalog of ETGs
that we restrict to a maximum apparent magnitude $17<i<\imax$ to
select the more massive systems in the redshift range more favorable
for lensing $( 0.2\lesssim z \lesssim 1 )$. We take advantage of a version of photometric redshift
measurements for the T06 CFHTLS data release updated by the work of
\citet{Cou++09} previously done for T04 earlier data release. These
were obtained using {\tt LePhare}
code\footnote{\url{http://www.cfht.hawaii.edu/$\sim$arnouts/LEPHARE/cfht\_lephare/lephare.html}}.
The reliability of photometric redshifts has been extensively assessed
against spectroscopic redshifts \citep{Ilb++06,Cou++09}. We define our
sample of ETGs as those galaxies having a best fit photometric
template Spectral Energy Distribution (SED) type $T<22$, corresponding to an SED
of an E/SO galaxy. The typical redshift uncertainty for these galaxies down
to $\imax$ is about $0.026$ with only 1.3\% of catastrophic
failures. We also exclude low redshift $z<0.1$ objects that are too
close for being efficient lenses and for which SED fitting leads to
greater redshift and type errors, given our filter coverage.

When looking for lenses, however, simple color cuts are not sufficient
to obtain complete samples of deflectors. A bright and blue strongly
lensed background source can alter significantly the colors of the
foreground deflector, misleading the classifier to think it is of a
different spectral type. In order not to miss some of the potentially
more promising targets we have thus augmented our sample of red
galaxies in the following fashion (A fully quantitative justification
of this procedure is given in Section~\ref{sssec:selcol} with the aid
of our extensive simulations).

\begin{figure}
\centering
\includegraphics[width=0.95\linewidth]{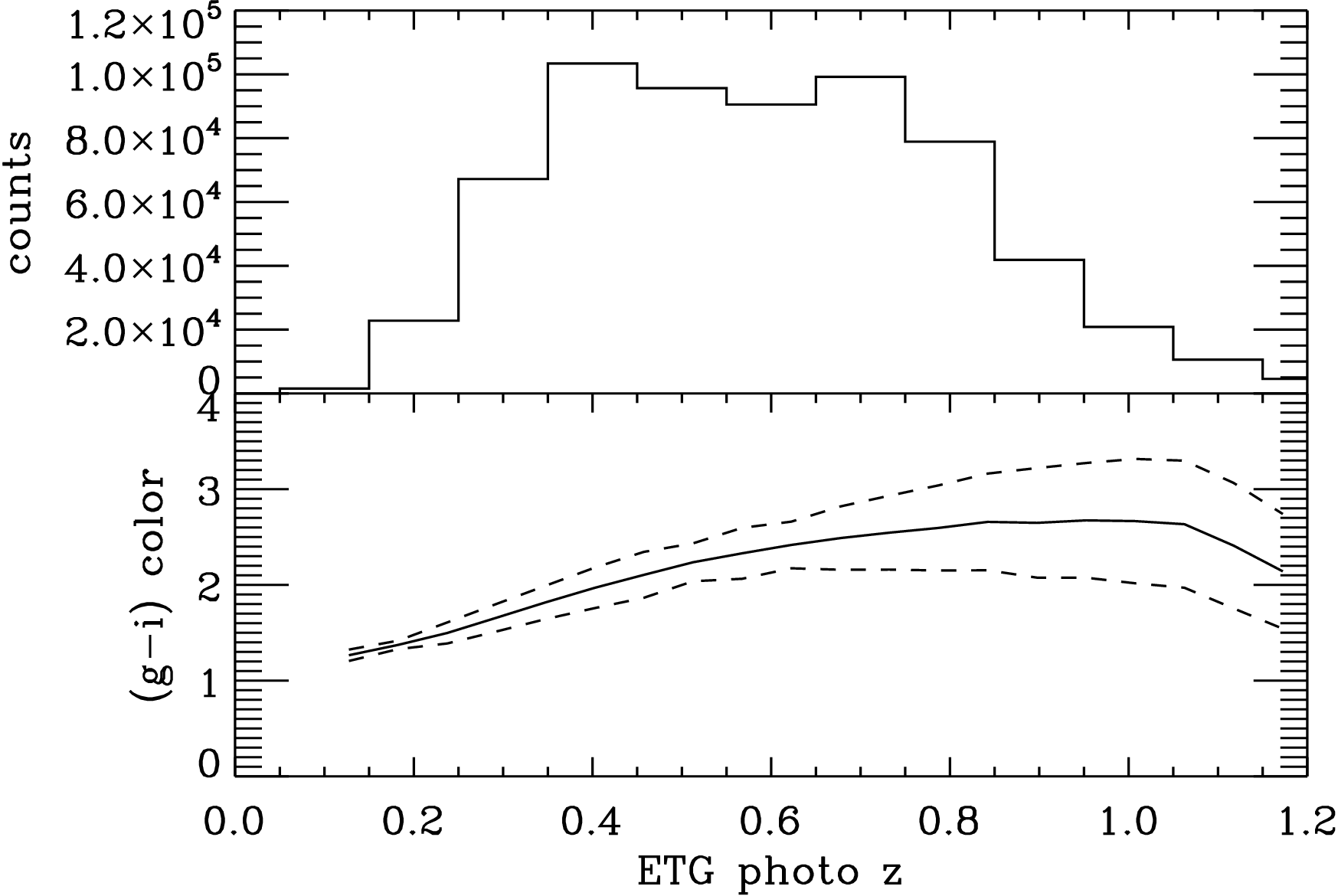}\\
\caption{\label{fig:ETGsel} {\it Top panel:} redshift distribution of Early-Type
Galaxies in the CFHTLS Wide down to a magnitude of $i<22$. {\it Bottom panel:}
mean (solid) and standard deviation (dashed) of the $\gmi_{\rm ETG}$ color
variation as a function of redshift.}
\end{figure}

We measure colors $\gmi_{\rm in}$ and $\gmi_{\rm out}$ in two
concentric circular apertures $R_{\rm in}=1\farcs86$ and $R_{\rm out}=3\farcs35$ and we
add to our catalog of potential lensing ETGs all galaxies that have a
red core and have a substantially bluer envelope regardless of their
best fit SED template type $T$. The criteria are as follow:
\begin{eqnarray}\label{eq:apercriterion}
\gmi_{\rm in} &>& 1.4 \label{eq:colsel1} \\ 
&{\rm and}\nonumber&\\ \gmi_{\rm in} - (g-i)_{\rm out} &<& 0.2
\;. \label{eq:colsel2}
\end{eqnarray}
As we show later these cuts are sufficient to add all the interesting
targets.  With this supplement, we typically end up with 3740 target ETGs per sq. degree (20\% of which have a
substantial color gradient). We are therefore left with $\sim
638\,000$ objects.  Their photometric redshift distribution along with
their $\gmi_{\rm ETG}$ color\footnote{Based on {\tt MAG\_AUTO}
photometry} as a function of redshift is shown in
Fig.~\ref{fig:ETGsel}. The median redshift is $z_{\rm med}=0.58$ with
16th and 84th percentiles of 0.36 and 0.82, respectively.  This
suggests that this parent dataset should be excellent for identifying
a sample of gravitational lenses with deflector redshift $z>0.4$, and
hence complement lower redshift samples like the SLACS.

\subsection{$g- \alpha\, i$ difference imaging}\label{sect:method:residuals}

For simplicity we approximate ETGs as having the same shape and radial
light profile in both $g$ and $i$ bands (we will justify this choice below).
 With this approximation,
lensed features that have different colors than the deflector will
show up in $g-\alpha\ i$ difference images, where $\alpha$ is an
appropriate scaling factor. As we will see, this approximation allows
us to remove the foreground deflector light robustly and with
sufficient precision to identify background lensed features as
positive residuals with blue colors.


\subsubsection{PSF matching and subtraction}\label{sect:method:residuals:PSF}

As a first step we have to match the spatial resolution of the red and
blue images. To achieve this, we simply convolve the red image with
the blue PSF, and vice-versa. Even though it entails some loss of
information, this process has the advantage of being robust provided
we have a good control of PSF variations over the large Megacam 1
deg$^2$ field-of-view. To understand the PSF, we build catalogs of
bright unsaturated stars in the magnitude range $17.3<i<21$ for each
CFHTLS pointing. For each ETG, we look for all the neighboring stars
less distant than 4 arcmin\footnote{Typically there are about 20 stars
within 4 arcmin in the high Galactic latitude W1, W2 and W3 fields,
and about 70 in the lower latitude W4 field.}, align them with
sub-pixel accuracy and make a flux-weighted average in each of the $g$
and $i$ bands. In most cases the image subtraction for ETGs leaves us
with very small residuals down to a radius of $\sim 0\farcs5$. On
smaller scales, imperfect PSF matching and/or some unaccounted-for
color gradient or nuclear emission prevents us from reaching residuals
consistent with noise. We thus conservatively discard residuals on
scales below $0\farcs6$. This effectively sets a lower limit to the
Einstein Radius of the gravitational lens systems that can be
identified by \RF.

The two light distributions are then matched by optimizing the value
of the scaling parameter $\alpha$ using a linear regression of the $g$ and
$i$ pixel values in the aperture $0\farcs5 \le R \le 2\arcsec$. We then
repeat a symmetric 3-sigma clipping of discrepant pixels four times to
end up with a $(g-i)$ color index map in which the deflector light is
suppressed.


\subsubsection{Deflector light profile analysis}\label{sect:method:residuals:analysis}

We use the sigma-clipped pixels described in the previous section to
perform a simple analysis of the red light distribution. In this
process we measure relevant quantities of the ETG, such as the second
order moments, axis ratio $q_{\rm L}$, orientation $\theta_{\rm l}$,
effective radius $R_{\rm eff,L}$ and S\'ersic index $n_{\rm L}$
\citep{Ser68}. Even though these quantities are affected by the PSF
blurring, they are useful to further refine the pre-selection of
ETGs by discarding spurious stars and red spirals.
In particular, by discarding the $\sim$12.5\% most
elongated objects with $q_{\rm L}<0.7$ and deflector having a S\'ersic index $n<1.5$,
we exclude disky foreground objects for which the assumption of an homogeneous color is
poor and will give rise to numerous false positives.


\subsubsection{Detection and analysis of residuals}\label{sect:method:residuals:resid}

In order to identify significant residuals, we measure the noise level
in the $g-\alpha\ i$ color map.  We then define a detection by
requiring a signal-to-noise-ratio $\nu>1.2$ over at least 10 connected
pixels in the unmasked image, provided that the center of light is in
the annulus $ R_{\rm min}=0\farcs5 \le R \le R_{\rm
max}=2\farcs7$. For each such connected region we infer the shape from
the zeroth, first and second order moments measured in the area
defined by the isophotal detection threshold.
In particular, for each connected region we
measure a flux $F_{\rm res}$, principal axes $a_{\rm res}$ and $b_{\rm
res}$ and their ratio $q_{\rm res}$, the radial separation $R_{\rm
res}$, and orientation $\varphi_{\rm res}$ of the main axis with
respect to the center of the deflector, and the isophotal area
$\mathcal{A}_{\rm res}$. We also keep track of the multiplicity
$\mathcal{M}$ of the residual, \ie~whether we detect several residuals
around a given foreground ETG.

Detections with width $b_{\rm res}<0\farcs2$ are considered spurious
given the typical image quality in both the $g$ and $i$ bands. After
this step, we find that about 14370 ETGs exhibit one or more
detectable blue residuals. Given the 135.2 deg$^2$ effective area,
this corresponds to about 106 objects per square degree.  We also
readily see that about 97.7\% of the parent population of ETGs
isolated in \S\ref{sect:method:selection} are automatically removed by
our automated \RF~pipeline.

However at this stage we have not yet taken advantage of the measured
quantities like $F_{\rm res}$, $q_{\rm res}$, $R_{\rm res}$ and
$\varphi_{\rm res}$ which as we will show will help further select
actual lens candidates and eliminate false positives.  As justified by
the simulations presented in
\S\ref{sect:simus} and the association of previously known lenses, we apply
the following cuts: 
\begin{itemize} 
\item the orientation of the major
axis of the residual should be nearly tangential with respect to the
center of the deflector. This reads $\vert \varphi\vert \le 30^\circ$.
\item the flux $F_{\rm res}$ should correspond to an AB
magnitude $m_{\rm res}\le 25.5$, the area $\mathcal{A}_{\rm res} < 7$
arcsec$^2$ and the mean surface brightness $< 26.3$ mag/arcsec$^2$.
\item If one and only one residual is found ($\mathcal{M}=1$), 
then we require it to be elongated $q_{\rm res}<0.7$. Otherwise, if $\mathcal{M}\ge2$, the
object is considered in the catalog of lens
candidates.
\end{itemize} 

After these cuts aiming at maximizing the recovery rate, we end up
having 2524 lens candidates passing all these automatic criteria,
corresponding to a spatial density of about 18 deg$^{-2}$.

\begin{figure*}
\centering
\includegraphics[width=0.99\linewidth]{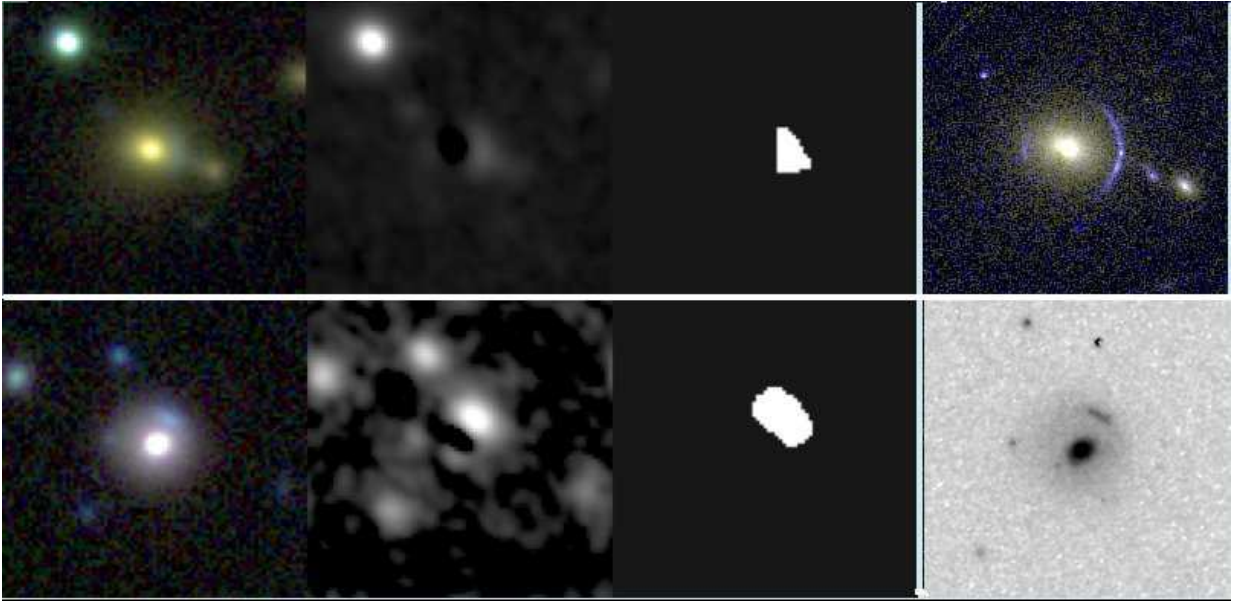}
\caption{\label{fig:RFexample} Two examples of \RF~outputs.
The first column is a {\it gri} composite color image.
The second column shows the result of the $g-\alpha i$ difference imaging
that performs well at suppressing the deflector light and picking the
blue arc-like residuals. The third column shows the detected
area(s) and the forth column shows the HST counterpart. Images are 15 arcsec on a side. In the first row we see a good candidate in a typical cusp configuration where only the bright outer image is detected. The second row illustrates a typical false positive consisting of a nearly face-on disky galaxy with a prominent bulge.}
\end{figure*}


\subsection{Visual inspection}\label{sect:method:inspection}

These 2524 candidates were subsequently visually scrutinized by the
authors in order to identify obvious spurious objects and refine our
lensing classification.  We thus defined a quality factor called {\tt q\_flag} taking
4 possible values, 0: not a lens, 1:
possibly a lens, 2: probably a lens, 3: definitely a lens. The classification is similar
to the scheme adopted for the SLACS survey \citep{Bol++04,Bol++06,Aug++09}. The visual
classification is very fast as the 2524 candidate lensing ETGs could
be visualized and classified in a few hours for a single
individual. When the eyeball classification of the authors disagreed,
a consensus classification for each object was achieved via a short
discussion.

The visual classification criteria are somewhat subjective as they
rely on the experience of the authors in observing, simulating, and
modeling confirmed lenses. It is an iterative process that builds on
the existence of some high resolution HST imaging of the COSMOS field
whose intersection with the CFHTLS deep field is one square degree and
in which a sample of gravitational lenses was published
\citep{Fau++08,Jac08}. In addition, as explained below, some early
follow up imaging with HST allowed us to train our classification.
Figures like in Fig.~\ref{fig:RFexample} were produced for known
lenses to see how they would look like and to know what to expect. It
is not trivial to figure out that the first row is an actual lens
whereas the second one is a nearly pole-on view of a disky galaxy with
a prominent bulge and a bright arm in the disk.
Note that \RF~yields very similar outputs for the three objects shown
in the figure. This illustrates that it will be difficult to
circumvent the need of a final informed visual inspection, or the need
for follow-up for final confirmation.

The visualization step allowed us to select \Nf~good and medium
quality lens candidates that are presented and further investigated in
\S\ref{sect:followup}. They are listed in table \ref{tab:cands}.

\renewcommand{\arraystretch}{1.10} 
\begin{deluxetable}{lrrcccccc}
\tablewidth{0pt}
\tabletypesize{\small}
\tablecaption{The SL2S galaxy-scale main sample of lens candidates in the CFHTLS-Wide.}
\tabletypesize{\footnotesize}
\tablehead{
\colhead{name SL2SJ...} &
\colhead{RA} &
\colhead{DEC} &
\colhead{mag$_i$} &
\colhead{$z_{\rm phot,d}$} & 
\colhead{$z_{\rm d}$} &
\colhead{$z_{\rm s}$} &
\colhead{\tt q\_flag} &
\colhead{\tt confirmed}
}
\startdata
020457$-$110309 &    31.2392 &   -11.0526 & 19.91 & 0.756 & 0.609 & 1.888 & 3 & 3 \\
020524$-$093023 &    31.3527 &    -9.5065 & 19.46 & 0.697 & 0.557 & 1.335 & 3 & 3 \\
020904$-$055529 &    32.2708 &    -5.9247 & 18.59 & 0.451 & \nodata & \nodata & 3 & \nodata \\
021206$-$075528 &    33.0273 &    -7.9245 & 19.04 & 0.494 & 0.460 & \nodata & 3 & 2 \\
021233$-$061210 &    33.1411 &    -6.2028 & 18.22 & 0.452 & \nodata & \nodata & 3 & \nodata \\
021247$-$055552 &    33.1993 &    -5.9312 & 20.46 & 0.806 & 0.750 & 2.740 & 3 & 3 \\
021411$-$040502 &    33.5467 &    -4.0841 & 19.88 & 0.740 & 0.609 & 1.880 & 3 & 3 \\
021517$-$061741 &    33.8222 &    -6.2948 & 18.06 & 0.388 & \nodata & \nodata & 3 & \nodata \\
021539$-$061918 &    33.9152 &    -6.3218 & 19.43 & 0.452 & \nodata & \nodata & 3 & \nodata \\
021737$-$051329 &    34.4049 &    -5.2248 & 19.63 & 0.856 & 0.646 & 1.850 & 3 & 3 \\
021801$-$080247 &    34.5053 &    -8.0465 & 20.45 & 1.058 & 0.928 & 2.060 & 3 & 3 \\
021902$-$082934 &    34.7589 &    -8.4930 & 18.98 & 0.499 & 0.389 & 2.160 & 3 & 3 \\
022046$-$094927 &    35.1919 &    -9.8244 & 19.93 & 0.603 & 0.572 & 2.606 & 3 & 3 \\
022056$-$063934 &    35.2358 &    -6.6595 & 18.08 & 0.389 & 0.330 & \nodata & 3 & 3 \\
022346$-$053418 &    35.9423 &    -5.5718 & 18.82 & 0.546 & 0.499 & 1.440 & 3 & 3 \\
022357$-$065142 &    35.9914 &    -6.8619 & 18.89 & 0.511 & 0.473 & 1.430 & 3 & 3 \\
022708$-$065445 &    36.7866 &    -6.9125 & 20.24 & 0.693 & 0.560 & 1.644 & 3 & 3 \\
023238$-$044948 &    38.1609 &    -4.8301 & 18.61 & 0.374 & \nodata & \nodata & 3 & \nodata \\
023307$-$043838 &    38.2794 &    -4.6440 & 19.62 & 0.786 & 0.671 & 1.869 & 3 & 3 \\
084934$-$043352 &   132.3926 &    -4.5646 & 18.59 & 0.395 & 0.373 & 2.400 & 3 & 2 \\
085019$-$034710 &   132.5795 &    -3.7863 & 19.50 & 0.333 & 0.337 & 3.250 & 3 & 3 \\
085503$-$023607 &   133.7648 &    -2.6020 & 20.39 & 0.679 & 0.622 & 0.351 & 3 & 0 \\
085540$-$014730 &   133.9172 &    -1.7918 & 19.57 & 0.425 & 0.365 & 3.390 & 3 & 3 \\
085559$-$040917 &   133.9996 &    -4.1549 & 18.90 & 0.450 & 0.419 & 2.950 & 3 & 3 \\
085831$-$035230 &   134.6317 &    -3.8751 & 19.60 & 0.750 & \nodata & \nodata & 3 & \nodata \\
135854+560349 &   209.7272 &    56.0639 & 18.88 & 0.594 & 0.499 & \nodata & 3 & \nodata \\
135949+553550 &   209.9567 &    55.5973 & 20.85 & 0.887 & 0.783 & 2.770 & 3 & 3 \\
140123+555705 &   210.3466 &    55.9515 & 19.06 & 0.639 & 0.527 & \nodata & 3 & 3 \\
140156+554446 &   210.4849 &    55.7463 & 18.70 & 0.504 & 0.464 & \nodata & 3 & 3 \\
140454+520024 &   211.2269 &    52.0068 & 17.88 & 0.490 & 0.456 & 1.590 & 3 & 3 \\
140533+550231 &   211.3910 &    55.0420 & 19.54 & 0.793 & \nodata & \nodata & 3 & 3 \\
140614+520253 &   211.5594 &    52.0482 & 18.54 & 0.510 & 0.480 & \nodata & 3 & 2 \\
140845+514913 &   212.1906 &    51.8205 & 19.65 & 0.739 & \nodata & \nodata & 3 & \nodata \\
141137+565119 &   212.9043 &    56.8553 & 18.74 & 0.415 & 0.322 & 1.420 & 3 & 3 \\
142059+563007 &   215.2494 &    56.5021 & 18.76 & 0.430 & 0.483 & 3.120 & 3 & 3 \\
142432+550019 &   216.1354 &    55.0055 & 19.39 & 0.451 & \nodata & \nodata & 3 & 0 \\
143341+512351 &   218.4209 &    51.3978 & 18.23 & 0.374 & \nodata & \nodata & 3 & \nodata \\
220329+020518 &   330.8709 &     2.0886 & 19.37 & 0.380 & 0.400 & 2.150 & 3 & 3 \\
221417+011855 &   333.5723 &     1.3154 & 19.99 & 0.510 & \nodata & \nodata & 3 & \nodata \\
221852+014038 &   334.7193 &     1.6775 & 19.24 & 0.612 & 0.564 & \nodata & 3 & 2 \\
222148+011542 &   335.4534 &     1.2619 & 18.35 & 0.346 & 0.325 & 2.350 & 3 & 3 \\
222217+001202 &   335.5735 &     0.2008 & 19.13 & 0.421 & 0.436 & 1.360 & 3 & 3 \\
020049$-$100048 &    30.2043 &   -10.0135 & 21.70 & 0.969 & \nodata & \nodata & 2 & \nodata \\
020103$-$034905 &    30.2666 &    -3.8181 & 20.54 & 0.724 & \nodata & \nodata & 2 & \nodata \\
020107$-$041845 &    30.2833 &    -4.3127 & 19.47 & 0.413 & \nodata & \nodata & 2 & \nodata \\
020148$-$084020 &    30.4510 &    -8.6723 & 19.61 & 0.476 & \nodata & \nodata & 2 & \nodata \\
020150$-$065235 &    30.4608 &    -6.8766 & 19.81 & 0.701 & \nodata & \nodata & 2 & \nodata \\
020150$-$103811 &    30.4608 &   -10.6366 & 18.78 & 0.529 & \nodata & \nodata & 2 & \nodata \\
020152$-$041103 &    30.4678 &    -4.1842 & 20.20 & 0.897 & \nodata & \nodata & 2 & \nodata \\
020201$-$063540 &    30.5049 &    -6.5945 & 18.81 & 0.418 & \nodata & \nodata & 2 & \nodata \\
020208$-$102006 &    30.5353 &   -10.3351 & 19.99 & 0.463 & \nodata & \nodata & 2 & \nodata \\
020232$-$071803 &    30.6355 &    -7.3009 & 19.84 & 0.463 & \nodata & \nodata & 2 & \nodata \\
020242$-$082113 &    30.6764 &    -8.3537 & 19.36 & 0.513 & \nodata & \nodata & 2 & \nodata \\
020308$-$100941 &    30.7845 &   -10.1616 & 19.71 & 0.368 & \nodata & \nodata & 2 & \nodata \\
020328$-$065719 &    30.8688 &    -6.9555 & 20.22 & 0.735 & \nodata & \nodata & 2 & \nodata \\
020338$-$051901 &    30.9097 &    -5.3171 & 18.77 & 0.432 & \nodata & \nodata & 2 & \nodata \\
020341$-$074722 &    30.9223 &    -7.7897 & 18.73 & 0.452 & \nodata & \nodata & 2 & \nodata \\
020342$-$035331 &    30.9269 &    -3.8920 & 18.79 & 0.289 & \nodata & \nodata & 2 & \nodata \\
020347$-$111201 &    30.9476 &   -11.2003 & 20.90 & 0.922 & \nodata & \nodata & 2 & \nodata \\
020353$-$100703 &    30.9739 &   -10.1176 & 20.37 & 0.562 & \nodata & \nodata & 2 & \nodata \\
020404$-$071418 &    31.0208 &    -7.2386 & 19.87 & 0.382 & \nodata & \nodata & 2 & \nodata \\
020408$-$061206 &    31.0368 &    -6.2019 & 19.92 & 0.440 & \nodata & \nodata & 2 & \nodata \\
020420$-$060940 &    31.0875 &    -6.1613 & 19.27 & 0.530 & \nodata & \nodata & 2 & \nodata \\
020425$-$060411 &    31.1071 &    -6.0699 & 20.56 & 0.789 & \nodata & \nodata & 2 & \nodata \\
020442$-$080650 &    31.1768 &    -8.1141 & 19.24 & 0.391 & \nodata & \nodata & 2 & \nodata \\
020451$-$100638 &    31.2131 &   -10.1106 & 19.28 & 0.414 & \nodata & \nodata & 2 & \nodata \\
020518$-$084524 &    31.3269 &    -8.7567 & 19.60 & 0.566 & \nodata & \nodata & 2 & \nodata \\
020523$-$080504 &    31.3460 &    -8.0847 & 19.80 & 0.429 & \nodata & \nodata & 2 & \nodata \\
020542$-$083423 &    31.4280 &    -8.5732 & 19.89 & 0.645 & \nodata & \nodata & 2 & \nodata \\
020601$-$043549 &    31.5054 &    -4.5972 & 18.63 & 0.389 & \nodata & \nodata & 2 & \nodata \\
020601$-$110253 &    31.5072 &   -11.0481 & 20.83 & 0.849 & \nodata & \nodata & 2 & \nodata \\
020603$-$094354 &    31.5153 &    -9.7317 & 19.95 & 0.679 & \nodata & \nodata & 2 & \nodata \\
020608$-$040928 &    31.5369 &    -4.1578 & 17.46 & 0.377 & \nodata & \nodata & 2 & \nodata \\
020646$-$042416 &    31.6926 &    -4.4046 & 19.76 & 0.634 & \nodata & \nodata & 2 & \nodata \\
020649$-$094957 &    31.7073 &    -9.8328 & 18.50 & 0.411 & \nodata & \nodata & 2 & \nodata \\
020651$-$075329 &    31.7162 &    -7.8916 & 20.73 & 0.792 & \nodata & \nodata & 2 & \nodata \\
020752$-$081759 &    31.9670 &    -8.2997 & 18.85 & 0.501 & \nodata & \nodata & 2 & \nodata \\
020817$-$073058 &    32.0709 &    -7.5162 & 18.84 & 0.518 & \nodata & \nodata & 2 & \nodata \\
020828$-$043254 &    32.1203 &    -4.5485 & 20.98 & 0.732 & \nodata & \nodata & 2 & \nodata \\
020828$-$045651 &    32.1184 &    -4.9476 & 18.17 & 0.329 & \nodata & \nodata & 2 & \nodata \\
020833$-$072434 &    32.1414 &    -7.4097 & 19.74 & 0.423 & \nodata & \nodata & 2 & \nodata \\
020848$-$073833 &    32.2016 &    -7.6427 & 20.16 & 0.688 & \nodata & \nodata & 2 & \nodata \\
020850$-$070217 &    32.2098 &    -7.0382 & 20.05 & 0.511 & \nodata & \nodata & 2 & \nodata \\
020855$-$090122 &    32.2311 &    -9.0228 & 18.88 & 0.446 & \nodata & \nodata & 2 & \nodata \\
020902$-$044814 &    32.2623 &    -4.8041 & 19.89 & 0.757 & \nodata & \nodata & 2 & \nodata \\
020909$-$063207 &    32.2899 &    -6.5354 & 18.16 & 0.480 & \nodata & \nodata & 2 & \nodata \\
020922$-$062506 &    32.3451 &    -6.4184 & 19.42 & 0.701 & \nodata & \nodata & 2 & \nodata \\
020925$-$072723 &    32.3577 &    -7.4565 & 20.84 & 0.536 & \nodata & \nodata & 2 & \nodata \\
021000$-$101256 &    32.5018 &   -10.2156 & 19.89 & 0.423 & \nodata & \nodata & 2 & \nodata \\
021001$-$074648 &    32.5065 &    -7.7800 & 18.25 & 0.420 & \nodata & \nodata & 2 & \nodata \\
021003$-$054006 &    32.5158 &    -5.6685 & 19.34 & 0.377 & \nodata & \nodata & 2 & \nodata \\
021025$-$043307 &    32.6056 &    -4.5522 & 20.68 & 0.651 & \nodata & \nodata & 2 & \nodata \\
021035$-$082837 &    32.6498 &    -8.4771 & 20.09 & 0.725 & \nodata & \nodata & 2 & \nodata \\
021040$-$074249 &    32.6706 &    -7.7138 & 20.35 & 0.708 & \nodata & \nodata & 2 & \nodata \\
021057$-$061238 &    32.7415 &    -6.2107 & 20.11 & 0.387 & \nodata & \nodata & 2 & \nodata \\
021101$-$085555 &    32.7568 &    -8.9320 & 20.67 & 0.562 & \nodata & \nodata & 2 & \nodata \\
021119$-$110308 &    32.8297 &   -11.0524 & 19.54 & 0.452 & \nodata & \nodata & 2 & \nodata \\
021121$-$041649 &    32.8405 &    -4.2803 & 20.61 & 0.397 & \nodata & \nodata & 2 & \nodata \\
021121$-$082353 &    32.8381 &    -8.3982 & 18.79 & 0.364 & \nodata & \nodata & 2 & \nodata \\
021122$-$104950 &    32.8438 &   -10.8308 & 19.26 & 0.508 & \nodata & \nodata & 2 & \nodata \\
021124$-$063951 &    32.8521 &    -6.6642 & 20.04 & 0.488 & \nodata & \nodata & 2 & \nodata \\
021155$-$075506 &    32.9801 &    -7.9185 & 19.76 & 0.375 & \nodata & \nodata & 2 & \nodata \\
021213$-$054849 &    33.0569 &    -5.8138 & 18.95 & 0.429 & \nodata & \nodata & 2 & \nodata \\
021222$-$042002 &    33.0948 &    -4.3340 & 20.76 & 0.528 & \nodata & \nodata & 2 & \nodata \\
021223$-$034530 &    33.0972 &    -3.7586 & 19.85 & 0.516 & \nodata & \nodata & 2 & \nodata \\
021228$-$074558 &    33.1180 &    -7.7664 & 20.06 & 0.736 & \nodata & \nodata & 2 & \nodata \\
021230$-$074727 &    33.1264 &    -7.7909 & 20.24 & 0.951 & \nodata & \nodata & 2 & \nodata \\
021237$-$091137 &    33.1569 &    -9.1939 & 19.39 & 0.459 & \nodata & \nodata & 2 & \nodata \\
021256$-$041555 &    33.2337 &    -4.2654 & 18.97 & 0.398 & \nodata & \nodata & 2 & \nodata \\
021303$-$064201 &    33.2644 &    -6.7005 & 19.21 & 0.429 & \nodata & \nodata & 2 & \nodata \\
021306$-$102026 &    33.2756 &   -10.3406 & 18.70 & 0.334 & \nodata & \nodata & 2 & \nodata \\
021323$-$065210 &    33.3496 &    -6.8696 & 19.64 & 0.568 & \nodata & \nodata & 2 & \nodata \\
021326$-$090618 &    33.3620 &    -9.1052 & 19.67 & 0.391 & \nodata & \nodata & 2 & \nodata \\
021429$-$070746 &    33.6247 &    -7.1295 & 20.75 & 0.786 & \nodata & \nodata & 2 & \nodata \\
021439$-$092631 &    33.6643 &    -9.4421 & 19.70 & 0.898 & \nodata & \nodata & 2 & \nodata \\
021535$-$080008 &    33.8968 &    -8.0024 & 19.17 & 0.447 & \nodata & \nodata & 2 & \nodata \\
021548$-$034752 &    33.9504 &    -3.7979 & 19.70 & 0.577 & \nodata & \nodata & 2 & \nodata \\
021613$-$061857 &    34.0582 &    -6.3161 & 19.82 & 0.824 & \nodata & \nodata & 2 & \nodata \\
021650$-$035948 &    34.2113 &    -3.9968 & 18.83 & 0.340 & \nodata & \nodata & 2 & \nodata \\
021714$-$081909 &    34.3096 &    -8.3194 & 20.47 & 0.396 & \nodata & \nodata & 2 & \nodata \\
021718$-$052921 &    34.3288 &    -5.4892 & 19.78 & 0.378 & \nodata & \nodata & 2 & \nodata \\
021725$-$085314 &    34.3578 &    -8.8872 & 19.01 & 0.481 & \nodata & \nodata & 2 & \nodata \\
021810$-$090954 &    34.5433 &    -9.1651 & 20.94 & 0.297 & \nodata & \nodata & 2 & \nodata \\
021823$-$053921 &    34.5983 &    -5.6559 & 19.85 & 0.814 & \nodata & \nodata & 2 & \nodata \\
021826$-$071727 &    34.6120 &    -7.2910 & 20.02 & 0.474 & \nodata & \nodata & 2 & \nodata \\
021829$-$060735 &    34.6228 &    -6.1266 & 20.61 & 0.903 & \nodata & \nodata & 2 & \nodata \\
021917$-$070239 &    34.8214 &    -7.0444 & 18.25 & 0.345 & \nodata & \nodata & 2 & \nodata \\
021923$-$053842 &    34.8470 &    -5.6451 & 21.66 & 0.902 & \nodata & \nodata & 2 & \nodata \\
021940$-$075456 &    34.9170 &    -7.9158 & 19.32 & 0.583 & \nodata & \nodata & 2 & \nodata \\
021950$-$055706 &    34.9596 &    -5.9518 & 19.66 & 0.769 & \nodata & \nodata & 2 & \nodata \\
021959$-$040607 &    34.9978 &    -4.1020 & 20.21 & 0.807 & \nodata & \nodata & 2 & \nodata \\
022005$-$035818 &    35.0221 &    -3.9719 & 18.42 & 0.449 & \nodata & \nodata & 2 & \nodata \\
022016$-$102446 &    35.0688 &   -10.4129 & 19.35 & 0.614 & \nodata & \nodata & 2 & \nodata \\
022046$-$100601 &    35.1927 &   -10.1005 & 21.44 & 0.694 & \nodata & \nodata & 2 & \nodata \\
022128$-$065953 &    35.3684 &    -6.9981 & 20.76 & 0.494 & \nodata & \nodata & 2 & \nodata \\
022133$-$075238 &    35.3899 &    -7.8774 & 19.46 & 0.372 & \nodata & \nodata & 2 & \nodata \\
022212$-$052610 &    35.5504 &    -5.4362 & 19.90 & 0.642 & \nodata & \nodata & 2 & \nodata \\
022245$-$100912 &    35.6883 &   -10.1535 & 19.83 & 0.747 & \nodata & \nodata & 2 & \nodata \\
022355$-$105518 &    35.9811 &   -10.9217 & 21.54 & 0.948 & \nodata & \nodata & 2 & \nodata \\
022359$-$073401 &    35.9971 &    -7.5671 & 20.21 & 0.614 & \nodata & \nodata & 2 & \nodata \\
022428$-$044422 &    36.1191 &    -4.7397 & 18.75 & 0.392 & \nodata & \nodata & 2 & \nodata \\
022458$-$050152 &    36.2421 &    -5.0312 & 18.29 & 0.405 & 0.361 & \nodata & 2 & \nodata \\
022527$-$035128 &    36.3644 &    -3.8579 & 19.23 & 0.388 & \nodata & \nodata & 2 & \nodata \\
022536$-$041517 &    36.4031 &    -4.2549 & 19.60 & 0.631 & 0.556 & \nodata & 2 & \nodata \\
022559$-$091844 &    36.4992 &    -9.3124 & 19.07 & 0.366 & \nodata & \nodata & 2 & \nodata \\
022603$-$094551 &    36.5152 &    -9.7643 & 18.30 & 0.229 & \nodata & \nodata & 2 & \nodata \\
022611$-$040646 &    36.5486 &    -4.1130 & 19.85 & 0.453 & \nodata & \nodata & 2 & \nodata \\
022612$-$072040 &    36.5508 &    -7.3447 & 20.02 & 0.476 & \nodata & \nodata & 2 & \nodata \\
022617$-$103728 &    36.5744 &   -10.6246 & 19.73 & 0.395 & \nodata & \nodata & 2 & \nodata \\
022633$-$034904 &    36.6385 &    -3.8180 & 20.08 & 0.652 & \nodata & \nodata & 2 & \nodata \\
022637$-$073627 &    36.6573 &    -7.6077 & 20.39 & 0.786 & \nodata & \nodata & 2 & \nodata \\
022658$-$080037 &    36.7455 &    -8.0105 & 19.06 & 0.450 & \nodata & \nodata & 2 & \nodata \\
022708$-$085753 &    36.7871 &    -8.9648 & 18.45 & 0.481 & \nodata & \nodata & 2 & \nodata \\
022817$-$080242 &    37.0727 &    -8.0452 & 19.57 & 0.437 & 0.483 & \nodata & 2 & \nodata \\
022820$-$094416 &    37.0851 &    -9.7378 & 19.34 & 0.475 & \nodata & \nodata & 2 & \nodata \\
022821$-$085203 &    37.0906 &    -8.8675 & 20.79 & 0.796 & \nodata & \nodata & 2 & \nodata \\
022834$-$074003 &    37.1435 &    -7.6677 & 20.62 & 1.016 & \nodata & \nodata & 2 & \nodata \\
022834$-$084314 &    37.1431 &    -8.7207 & 19.08 & 0.493 & \nodata & \nodata & 2 & \nodata \\
022841$-$061729 &    37.1721 &    -6.2914 & 20.32 & 0.700 & \nodata & \nodata & 2 & \nodata \\
022914$-$080515 &    37.3097 &    -8.0877 & 19.97 & 0.752 & \nodata & \nodata & 2 & \nodata \\
022915$-$060902 &    37.3153 &    -6.1507 & 20.14 & 0.732 & \nodata & \nodata & 2 & \nodata \\
022923$-$104425 &    37.3498 &   -10.7403 & 18.68 & 0.337 & \nodata & \nodata & 2 & \nodata \\
022942$-$041529 &    37.4263 &    -4.2582 & 20.97 & 1.002 & \nodata & \nodata & 2 & \nodata \\
023010$-$110409 &    37.5420 &   -11.0694 & 20.53 & 0.896 & \nodata & \nodata & 2 & \nodata \\
023026$-$044654 &    37.6092 &    -4.7819 & 20.64 & 0.698 & \nodata & \nodata & 2 & \nodata \\
023031$-$065103 &    37.6307 &    -6.8510 & 19.75 & 0.660 & \nodata & \nodata & 2 & \nodata \\
023047$-$110210 &    37.6981 &   -11.0363 & 21.07 & 1.002 & \nodata & \nodata & 2 & \nodata \\
023049$-$094140 &    37.7056 &    -9.6946 & 20.28 & 0.703 & \nodata & \nodata & 2 & \nodata \\
023134$-$044922 &    37.8937 &    -4.8229 & 18.18 & 0.420 & 0.393 & \nodata & 2 & \nodata \\
023145$-$061313 &    37.9407 &    -6.2205 & 18.54 & 0.410 & \nodata & \nodata & 2 & \nodata \\
023148$-$100603 &    37.9501 &   -10.1011 & 20.03 & 0.766 & \nodata & \nodata & 2 & \nodata \\
023150$-$041730 &    37.9618 &    -4.2917 & 19.69 & 0.838 & \nodata & \nodata & 2 & \nodata \\
023152$-$103008 &    37.9685 &   -10.5025 & 20.84 & 0.342 & \nodata & \nodata & 2 & \nodata \\
023248$-$063321 &    38.2014 &    -6.5560 & 19.29 & 0.461 & \nodata & \nodata & 2 & \nodata \\
023253$-$083436 &    38.2216 &    -8.5769 & 18.07 & 0.359 & \nodata & \nodata & 2 & \nodata \\
023255$-$062123 &    38.2318 &    -6.3564 & 18.83 & 0.430 & \nodata & \nodata & 2 & 1 \\
023322$-$055202 &    38.3417 &    -5.8674 & 18.12 & 0.462 & 0.434 & \nodata & 2 & \nodata \\
023325$-$053104 &    38.3547 &    -5.5178 & 18.72 & 0.467 & \nodata & \nodata & 2 & \nodata \\
023431$-$095636 &    38.6329 &    -9.9435 & 21.12 & 0.759 & \nodata & \nodata & 2 & \nodata \\
023444$-$064832 &    38.6843 &    -6.8091 & 20.32 & 0.728 & \nodata & \nodata & 2 & \nodata \\
023511$-$051752 &    38.7968 &    -5.2978 & 20.32 & 0.420 & \nodata & \nodata & 2 & \nodata \\
084838$-$035319 &   132.1600 &    -3.8887 & 20.94 & 0.738 & \nodata & \nodata & 2 & \nodata \\
084847$-$035103 &   132.1968 &    -3.8511 & 20.75 & 0.885 & 0.682 & 1.550 & 2 & 2 \\
084921$-$024531 &   132.3383 &    -2.7588 & 19.97 & 0.514 & \nodata & \nodata & 2 & \nodata \\
084924$-$031521 &   132.3540 &    -3.2559 & 19.85 & 0.335 & \nodata & \nodata & 2 & \nodata \\
084941$-$051650 &   132.4216 &    -5.2808 & 19.06 & 0.392 & \nodata & \nodata & 2 & \nodata \\
084959$-$025142 &   132.4990 &    -2.8619 & 18.19 & 0.305 & 0.274 & 2.090 & 2 & 3 \\
085009$-$024703 &   132.5397 &    -2.7842 & 18.80 & 0.437 & \nodata & \nodata & 2 & \nodata \\
085018$-$023240 &   132.5784 &    -2.5447 & 20.11 & 0.731 & \nodata & \nodata & 2 & \nodata \\
085039$-$025458 &   132.6658 &    -2.9162 & 21.09 & 0.717 & \nodata & \nodata & 2 & \nodata \\
085046$-$041458 &   132.6918 &    -4.2496 & 19.74 & 0.393 & \nodata & \nodata & 2 & \nodata \\
085135$-$041456 &   132.8989 &    -4.2490 & 20.21 & 0.648 & \nodata & \nodata & 2 & \nodata \\
085203$-$040111 &   133.0161 &    -4.0198 & 20.05 & 0.636 & \nodata & \nodata & 2 & \nodata \\
085233$-$051502 &   133.1381 &    -5.2506 & 19.92 & 0.696 & \nodata & \nodata & 2 & \nodata \\
085317$-$020312 &   133.3233 &    -2.0535 & 20.51 & 0.706 & \nodata & \nodata & 2 & \nodata \\
085327$-$023745 &   133.3655 &    -2.6292 & 20.33 & 0.109 & 0.774 & 2.440 & 2 & 2 \\
085508$-$030607 &   133.7865 &    -3.1020 & 20.64 & 0.613 & \nodata & \nodata & 2 & \nodata \\
085713$-$043809 &   134.3052 &    -4.6360 & 19.53 & 0.491 & \nodata & \nodata & 2 & \nodata \\
085719$-$023807 &   134.3328 &    -2.6355 & 20.33 & 0.971 & \nodata & \nodata & 2 & \nodata \\
085749$-$023455 &   134.4549 &    -2.5821 & 20.67 & 0.720 & \nodata & \nodata & 2 & \nodata \\
085816$-$030954 &   134.5704 &    -3.1652 & 20.56 & 0.872 & \nodata & \nodata & 2 & \nodata \\
085907$-$042147 &   134.7800 &    -4.3631 & 19.55 & 0.379 & \nodata & \nodata & 2 & \nodata \\
085912$-$032248 &   134.8003 &    -3.3801 & 21.09 & 0.907 & \nodata & \nodata & 2 & \nodata \\
085953$-$041754 &   134.9723 &    -4.2986 & 19.88 & 0.436 & \nodata & \nodata & 2 & \nodata \\
090019$-$014745 &   135.0805 &    -1.7961 & 19.04 & 0.546 & \nodata & \nodata & 2 & \nodata \\
090036$-$051944 &   135.1506 &    -5.3290 & 20.64 & 0.711 & \nodata & \nodata & 2 & \nodata \\
090154$-$034046 &   135.4761 &    -3.6795 & 18.73 & 0.484 & \nodata & \nodata & 2 & \nodata \\
090216$-$014057 &   135.5702 &    -1.6828 & 18.44 & 0.401 & \nodata & \nodata & 2 & \nodata \\
090217$-$015130 &   135.5743 &    -1.8585 & 20.07 & 0.392 & \nodata & \nodata & 2 & \nodata \\
090604$-$035611 &   136.5195 &    -3.9365 & 19.51 & 0.776 & \nodata & \nodata & 2 & \nodata \\
090630$-$043236 &   136.6269 &    -4.5435 & 19.32 & 0.512 & \nodata & \nodata & 2 & \nodata \\
090650$-$033108 &   136.7089 &    -3.5191 & 20.79 & 0.949 & \nodata & \nodata & 2 & \nodata \\
090706$-$013055 &   136.7761 &    -1.5153 & 18.87 & 0.430 & \nodata & \nodata & 2 & \nodata \\
135503+564447 &   208.7649 &    56.7465 & 20.72 & 0.803 & \nodata & \nodata & 2 & \nodata \\
135552+555806 &   208.9682 &    55.9684 & 19.31 & 0.452 & \nodata & \nodata & 2 & \nodata \\
135634+552912 &   209.1426 &    55.4869 & 19.59 & 0.575 & \nodata & \nodata & 2 & \nodata \\
135655+573550 &   209.2315 &    57.5972 & 20.65 & 0.927 & \nodata & \nodata & 2 & \nodata \\
135738+554822 &   209.4110 &    55.8061 & 19.70 & 0.494 & \nodata & \nodata & 2 & \nodata \\
135804+551506 &   209.5202 &    55.2517 & 19.63 & 0.422 & \nodata & \nodata & 2 & \nodata \\
135851+563515 &   209.7126 &    56.5876 & 20.15 & 0.552 & \nodata & \nodata & 2 & \nodata \\
140021+513122 &   210.0897 &    51.5230 & 19.82 & 0.523 & \nodata & \nodata & 2 & \nodata \\
140022+545804 &   210.0947 &    54.9680 & 20.01 & 0.703 & \nodata & \nodata & 2 & \nodata \\
140042+560042 &   210.1775 &    56.0118 & 19.26 & 0.568 & \nodata & \nodata & 2 & \nodata \\
140139+522037 &   210.4125 &    52.3437 & 20.63 & 0.668 & \nodata & \nodata & 2 & \nodata \\
140212+574516 &   210.5509 &    57.7547 & 19.46 & 0.458 & \nodata & \nodata & 2 & \nodata \\
140221+550534 &   210.5897 &    55.0930 & 18.44 & 0.454 & 0.412 & \nodata & 2 & 3 \\
140222+541003 &   210.5932 &    54.1676 & 19.31 & 0.336 & \nodata & \nodata & 2 & \nodata \\
140225+563946 &   210.6061 &    56.6629 & 20.32 & 0.662 & \nodata & \nodata & 2 & \nodata \\
140254+550516 &   210.7288 &    55.0878 & 19.11 & 0.354 & \nodata & \nodata & 2 & \nodata \\
140254+551324 &   210.7255 &    55.2234 & 19.80 & 0.541 & \nodata & \nodata & 2 & \nodata \\
140257+520712 &   210.7412 &    52.1201 & 20.28 & 0.647 & \nodata & \nodata & 2 & \nodata \\
140313+543839 &   210.8060 &    54.6442 & 18.61 & 0.375 & \nodata & \nodata & 2 & \nodata \\
140340+555229 &   210.9207 &    55.8749 & 20.76 & 0.771 & \nodata & \nodata & 2 & \nodata \\
140340+564607 &   210.9173 &    56.7688 & 19.54 & 0.689 & \nodata & \nodata & 2 & \nodata \\
140414+555205 &   211.0606 &    55.8681 & 19.89 & 0.737 & \nodata & \nodata & 2 & \nodata \\
140424+513845 &   211.1027 &    51.6459 & 19.16 & 0.641 & \nodata & \nodata & 2 & \nodata \\
140458+522549 &   211.2449 &    52.4305 & 18.59 & 0.400 & \nodata & \nodata & 2 & \nodata \\
140503+555441 &   211.2651 &    55.9117 & 20.03 & 0.710 & \nodata & \nodata & 2 & \nodata \\
140517+543549 &   211.3248 &    54.5971 & 20.48 & 0.726 & \nodata & \nodata & 2 & \nodata \\
140523+512403 &   211.3469 &    51.4009 & 19.89 & 0.605 & \nodata & \nodata & 2 & \nodata \\
140535+571811 &   211.3971 &    57.3032 & 19.08 & 0.652 & \nodata & \nodata & 2 & \nodata \\
140546+524311 &   211.4426 &    52.7198 & 19.26 & 0.546 & 0.526 & \nodata & 2 & 3 \\
140635+542325 &   211.6471 &    54.3904 & 21.12 & 0.817 & \nodata & \nodata & 2 & \nodata \\
140717+513522 &   211.8220 &    51.5896 & 18.44 & 0.367 & \nodata & \nodata & 2 & \nodata \\
140732+543408 &   211.8857 &    54.5690 & 19.03 & 0.411 & \nodata & \nodata & 2 & \nodata \\
140751+550230 &   211.9640 &    55.0417 & 19.54 & 0.359 & \nodata & \nodata & 2 & \nodata \\
140855+524452 &   212.2298 &    52.7479 & 19.98 & 0.492 & \nodata & \nodata & 2 & \nodata \\
140910+544645 &   212.2918 &    54.7794 & 20.30 & 0.699 & \nodata & \nodata & 2 & \nodata \\
140935+541711 &   212.3998 &    54.2864 & 19.75 & 0.837 & \nodata & \nodata & 2 & \nodata \\
140949+524818 &   212.4549 &    52.8050 & 19.64 & 0.437 & \nodata & \nodata & 2 & 1 \\
141017+535335 &   212.5729 &    53.8932 & 20.29 & 0.731 & \nodata & \nodata & 2 & \nodata \\
141023+531635 &   212.5969 &    53.2766 & 19.26 & 0.394 & \nodata & \nodata & 2 & \nodata \\
141056+533225 &   212.7364 &    53.5405 & 19.33 & 0.717 & \nodata & \nodata & 2 & \nodata \\
141139+535339 &   212.9133 &    53.8944 & 20.69 & 0.458 & \nodata & \nodata & 2 & \nodata \\
141206+535059 &   213.0268 &    53.8498 & 18.07 & 0.450 & 0.391 & \nodata & 2 & \nodata \\
141211+574514 &   213.0495 &    57.7541 & 20.52 & 0.767 & \nodata & \nodata & 2 & \nodata \\
141216+534223 &   213.0681 &    53.7064 & 19.44 & 0.591 & \nodata & \nodata & 2 & \nodata \\
141228+531220 &   213.1167 &    53.2056 & 18.99 & 0.556 & \nodata & \nodata & 2 & \nodata \\
141257+530120 &   213.2380 &    53.0223 & 20.03 & 0.419 & \nodata & \nodata & 2 & \nodata \\
141504+522823 &   213.7694 &    52.4733 & 19.43 & 0.510 & \nodata & \nodata & 2 & \nodata \\
141543+522735 &   213.9302 &    52.4597 & 19.17 & 0.445 & \nodata & \nodata & 2 & \nodata \\
142027+540842 &   215.1154 &    54.1452 & 18.54 & 0.421 & \nodata & \nodata & 2 & \nodata \\
142031+525822 &   215.1325 &    52.9728 & 18.87 & 0.461 & 0.380 & 0.990 & 2 & 2 \\
142044+544900 &   215.1850 &    54.8167 & 19.97 & 0.727 & \nodata & \nodata & 2 & \nodata \\
142119+531109 &   215.3296 &    53.1859 & 19.44 & 0.617 & \nodata & \nodata & 2 & \nodata \\
142254+564909 &   215.7268 &    56.8192 & 20.30 & 0.740 & \nodata & \nodata & 2 & \nodata \\
142258+512439 &   215.7430 &    51.4110 & 20.70 & 0.736 & \nodata & \nodata & 2 & \nodata \\
142311+513926 &   215.7972 &    51.6574 & 20.64 & 0.888 & \nodata & \nodata & 2 & \nodata \\
142423+523353 &   216.0988 &    52.5648 & 18.31 & 0.277 & \nodata & \nodata & 2 & \nodata \\
142501+514652 &   216.2574 &    51.7814 & 18.35 & 0.396 & \nodata & \nodata & 2 & \nodata \\
142506+525206 &   216.2764 &    52.8684 & 19.43 & 0.633 & \nodata & \nodata & 2 & \nodata \\
142731+551645 &   216.8803 &    55.2792 & 19.98 & 0.587 & 0.410 & 2.580 & 2 & 3 \\
142732+554230 &   216.8833 &    55.7084 & 17.66 & 0.180 & \nodata & \nodata & 2 & \nodata \\
142740+555127 &   216.9184 &    55.8578 & 17.95 & 0.366 & \nodata & \nodata & 2 & \nodata \\
142827+522458 &   217.1156 &    52.4161 & 19.42 & 0.456 & \nodata & \nodata & 2 & \nodata \\
142834+552736 &   217.1436 &    55.4600 & 20.32 & 0.741 & \nodata & \nodata & 2 & \nodata \\
142858+521606 &   217.2422 &    52.2684 & 19.07 & 0.531 & \nodata & \nodata & 2 & \nodata \\
142943+545330 &   217.4292 &    54.8918 & 18.15 & 0.310 & \nodata & \nodata & 2 & \nodata \\
142948+543013 &   217.4506 &    54.5037 & 18.91 & 0.451 & \nodata & \nodata & 2 & \nodata \\
143001+554334 &   217.5065 &    55.7264 & 19.88 & 0.689 & \nodata & \nodata & 2 & \nodata \\
143013+550052 &   217.5583 &    55.0145 & 19.45 & 0.653 & \nodata & \nodata & 2 & \nodata \\
143101+541320 &   217.7573 &    54.2225 & 20.09 & 0.931 & \nodata & \nodata & 2 & \nodata \\
143123+544819 &   217.8494 &    54.8055 & 20.45 & 0.963 & \nodata & \nodata & 2 & \nodata \\
143130+570931 &   217.8789 &    57.1586 & 18.71 & 0.454 & \nodata & \nodata & 2 & \nodata \\
143151+515531 &   217.9641 &    51.9255 & 19.68 & 0.488 & \nodata & \nodata & 2 & \nodata \\
143157+524645 &   217.9876 &    52.7793 & 18.93 & 0.495 & \nodata & \nodata & 2 & \nodata \\
143208+534419 &   218.0356 &    53.7388 & 19.10 & 0.612 & \nodata & \nodata & 2 & \nodata \\
143341+524150 &   218.4230 &    52.6973 & 20.10 & 0.493 & \nodata & \nodata & 2 & \nodata \\
143421+543814 &   218.5875 &    54.6375 & 20.29 & 0.728 & \nodata & \nodata & 2 & \nodata \\
143436+531743 &   218.6511 &    53.2955 & 18.84 & 0.511 & \nodata & \nodata & 2 & \nodata \\
143457+570936 &   218.7415 &    57.1602 & 20.79 & 0.335 & \nodata & \nodata & 2 & \nodata \\
143622+572740 &   219.0926 &    57.4613 & 20.27 & 0.652 & \nodata & \nodata & 2 & \nodata \\
143637+545636 &   219.1556 &    54.9436 & 19.05 & 0.360 & \nodata & \nodata & 2 & \nodata \\
143719+573714 &   219.3332 &    57.6208 & 19.40 & 0.537 & \nodata & \nodata & 2 & \nodata \\
143727+562144 &   219.3666 &    56.3624 & 21.21 & 0.794 & \nodata & \nodata & 2 & \nodata \\
143906+543900 &   219.7768 &    54.6502 & 19.33 & 0.378 & \nodata & \nodata & 2 & \nodata \\
143908+545250 &   219.7837 &    54.8808 & 18.93 & 0.543 & \nodata & \nodata & 2 & \nodata \\
220231+042458 &   330.6301 &     4.4163 & 18.05 & 0.348 & \nodata & \nodata & 2 & \nodata \\
220241+012612 &   330.6709 &     1.4369 & 18.12 & 0.339 & \nodata & \nodata & 2 & \nodata \\
220252+021336 &   330.7197 &     2.2269 & 20.18 & 0.305 & \nodata & \nodata & 2 & \nodata \\
220259+033640 &   330.7463 &     3.6112 & 19.02 & 0.485 & \nodata & \nodata & 2 & \nodata \\
220331+040310 &   330.8819 &     4.0530 & 21.15 & 0.632 & \nodata & \nodata & 2 & \nodata \\
220506+014703 &   331.2788 &     1.7844 & 19.15 & 0.460 & 0.476 & 2.520 & 2 & 3 \\
220604+014048 &   331.5168 &     1.6802 & 20.73 & 0.946 & 0.874 & \nodata & 2 & \nodata \\
220629+005728 &   331.6225 &     0.9580 & 19.77 & 0.759 & 0.704 & \nodata & 2 & 3 \\
220722+013610 &   331.8441 &     1.6030 & 20.09 & 0.823 & \nodata & \nodata & 2 & \nodata \\
220732+031311 &   331.8869 &     3.2199 & 19.84 & 0.831 & 0.663 & \nodata & 2 & \nodata \\
220759+002157 &   331.9962 &     0.3661 & 19.03 & 0.291 & \nodata & \nodata & 2 & \nodata \\
220838+030108 &   332.1595 &     3.0189 & 18.51 & 0.302 & \nodata & \nodata & 2 & \nodata \\
220851+004622 &   332.2137 &     0.7729 & 19.32 & 0.535 & \nodata & \nodata & 2 & \nodata \\
221000$-$000041 &   332.5023 &    -0.0116 & 18.53 & 0.389 & \nodata & \nodata & 2 & \nodata \\
221101+003401 &   332.7582 &     0.5671 & 19.97 & 0.895 & 0.676 & \nodata & 2 & \nodata \\
221225+000403 &   333.1049 &     0.0676 & 19.03 & 0.470 & \nodata & \nodata & 2 & \nodata \\
221236+014816 &   333.1500 &     1.8046 & 19.02 & 0.503 & \nodata & \nodata & 2 & \nodata \\
221238$-$001727 &   333.1587 &    -0.2910 & 18.70 & 0.428 & \nodata & \nodata & 2 & \nodata \\
221329+002935 &   333.3724 &     0.4932 & 18.91 & 0.483 & \nodata & \nodata & 2 & \nodata \\
221336+001143 &   333.4012 &     0.1954 & 20.34 & 0.623 & \nodata & \nodata & 2 & \nodata \\
221359+005416 &   333.4959 &     0.9046 & 18.27 & 0.370 & \nodata & \nodata & 2 & \nodata \\
221455+012932 &   333.7316 &     1.4923 & 19.85 & 0.768 & 0.591 & \nodata & 2 & \nodata \\
221457+010228 &   333.7383 &     1.0413 & 19.07 & 0.490 & \nodata & \nodata & 2 & \nodata \\
221519+015748 &   333.8305 &     1.9635 & 19.33 & 0.410 & \nodata & \nodata & 2 & \nodata \\
221627+021207 &   334.1165 &     2.2020 & 19.15 & 0.482 & \nodata & \nodata & 2 & \nodata \\
221649+021529 &   334.2071 &     2.2581 & 17.87 & 0.330 & 0.260 & \nodata & 2 & \nodata \\
221731+020715 &   334.3812 &     2.1210 & 20.52 & 0.846 & \nodata & \nodata & 2 & \nodata \\
221821+021557 &   334.5901 &     2.2660 & 19.56 & 0.492 & \nodata & \nodata & 2 & \nodata \\
221929$-$001743 &   334.8725 &    -0.2954 & 17.89 & 0.296 & 0.289 & 1.020 & 2 & 3 \\
222007$-$002505 &   335.0307 &    -0.4182 & 21.29 & 0.342 & 0.911 & \nodata & 2 & \nodata \\
222012+010606 &   335.0536 &     1.1018 & 18.83 & 0.240 & 0.232 & 1.070 & 2 & 2 \\
222131+012306 &   335.3807 &     1.3853 & 17.70 & 0.393 & 0.333 & \nodata & 2 & \nodata \\
222240+010951 &   335.6696 &     1.1642 & 18.95 & 0.415 & \nodata & \nodata & 2 & \nodata \\

\enddata
\tablecomments{\label{tab:cands} Photometric redshifts were measured by \citet{Cou++09}. Deflector $\zd$ and source $\zs$ redshifts, when measured, as listed. The follow-up confirmation flags {\tt confirmed} are also listed when some additional dataset brought firmer pieces of evidence on the nature of the candidate previously classified as either a good candidate {\tt q\_flag=2} or an excellent candidate {\tt q\_flag=3}. mag$_i$ refers to the $i$ band apparent magnitude of the candidate deflector. Systems are sorted in ascending name order with a first block of {\tt q\_flag=3} values first, followed by a block of {\tt q\_flag=2}.}
\end{deluxetable}

In addition to this clearly defined sample of lens candidates, we
present in Table~\ref{tab:cands2}, \NfX~systems that were detected with
earlier implementations of \RF~or previous CFHTLS data releases.
Some of the systems
presented were serendipitously found in the CFHTLS Deep survey. This
sample is not meant as a statistical sample and therefore we have not
carried out a complete statistical analysis of its selection
function. Candidates are shown anyway as some of them were considered
for follow-up imaging or spectroscopy and they might be useful for
future work.


\section{Simulating CFHTLS lenses}
\label{sect:simus}  

In order to understand the efficiency of \RF at recovering actual
lenses (completeness), and the fraction of true lenses among all the
detections (purity), we need a validation set of known lenses whose
mock observables (images) have been run through the \RF~detection
pipeline. Since we do not have a sufficiently large sample of real
gravitational lenses, we use realistic simulations instead.  In this
section, we first describe the physical assumptions that we put in to 
our simulated lenses, realize a sample of mock lenses as they could
appear in the CFHTLS Wide data, and then feed them through the
\RF~pipeline.


\subsection{The background source population}

Our lens survey is essentially surface brightness limited in the $g$
band\footnote{More precisely, we are surface brightness limited in the
complex $(g-\alpha i)$ difference images
(\S\ref{sect:method:residuals})}. We thus need to consider, in a
self-consistent way, the multivariate distribution of redshift $\zs$,
half-light radius $\rsrc$, $i$ band magnitude and $\gmi$ color for the
population of background sources that might be strongly lensed. We use
the COSMOS30 catalogs from the deep COSMOS survey \citep{Ilb++09} to
account for any magnification bias that might be introduced to the
lensed sources, but also to obtain approximate pre-seeing galaxy
sizes. Therefore, instead of drawing multivariate realizations of
faint background sources in a Monte-Carlo approach from an analytic
expression, we simply draw randomly a background source from the
COSMOS catalog and consider its full set of $i$, $\gmi$, $z$  and
$\rsrc$ values. The COSMOS catalog is complete down to $i\sim25$ and
takes advantage of 30 broad and narrow band filters covering UV to
mid-IR wavelengths. The space density of such a population is $n_{\rm
bg} \simeq 40\, {\rm arcmin}^{-2}$.

We choose to model all our background sources with exponential
elliptical profiles, with ellipticity~$e$ drawn from a Rayleigh
distribution of dispersion $\sigma_e=0.3$ \citep[as is often used in 
weak lensing studies, e.g.][and references therein]{Mil++13} but
limited to $e_{\rm max}=0.8$. The sources' orientations are assumed to
be uniformly distributed between 0 and $\pi$ radians.

 
\subsection{Population of foreground deflectors}

A key feature of lens-oriented searches is that they require a
pre-selection of potential deflectors. A magnitude-limited sample of
bright $i<\imax$ ETGs will readily lead to a selection of galaxies of
significant mass, and therefore lens strength, but whose completeness
that will vary with redshift.

We assume that the deflectors can be modeled as Singular Isothermal
Ellipsoids (SIE) with a characteristic velocity dispersion $\sigma$.
This latter quantity and the effective radius $\Reff$ of the deflector
can be uniquely related to its $i$ band absolute magnitude $M_i$
through the Fundamental Plane \citep{Dre++87,D+D87}.  We first assign
$\Reff$ using the Kormendy relation \citep{Kor77}:
\begin{equation} \label{eq:scaleKor}
	\logt \frac{\Reff}{\kpc} =  0.62 + 0.26667 \times (-21.64 - M_i) + \varepsilon_{\Reff}\;.
\end{equation}
where $\varepsilon_{\Reff}$ captures the scatter in that relation and
we assume a value of $0.11$ (\ie~30\% intrinsic scatter).  We then
estimate the mean effective surface density $\mu_0$ and use the recent
values of \citet{Ber++05} and \citet{HydeBernardi2009}:  $a=1.404$,
$b=0.304$ and $c=-8.858$ to get $\sigma$ such that:
\begin{equation} \label{eq:scaleFJ}
	a \logt \frac{\sigma}{\kms} = \logt \frac{\Reff}{\kpc} - b \mu_0 -c 
\end{equation}
The values are consistent with the assumptions made by \citet{oguri06}
to calculate lensing optical depths. We neglect the redshift
dependencies of these relations.
In practice, the sample of simulated ETGs has a median velocity dispersion of
210 $\kms$ and an rms dispersion of about $65 \kms$, with a mild shift of the 
distribution toward higher (resp. lower) values with redshift, which is an obvious
translation of the selection in apparent magnitude.

Despite this model's simplicity, the inner parts of all massive ETGs
have been found to be well approximated by  Singular Isothermal
Ellipsoids (SIE) \citep[see e.g.][]{RKK03,Koo++06,Koo++09}. We can thus
define the Einstein radius $R_{\rm Ein}$ as:
\begin{equation}
   R_{\rm Ein} = 4 \pi \left( \frac{\sigma}{c}\right)^2  w \;{\rm rd}
	                   = \left(\frac{\sigma}{186.21\, \kms}\right)^2 w\;{\rm arcsec}\;,
\end{equation}
where $w\equiv d_{\rm ls} / d_{\rm s} $ is the ratio of distances
between the deflector and the source, and between the observer and the
source.

The ellipticity and orientation of the elliptical total mass
distribution are assumed to be that of the deflector's stellar light.
Neglecting the presence of either intrinsic misalignment or external
shear in this way is justified by the super-critical parts of the
galaxy density distribution being dominated by stellar mass, and by
the fact that the lens statistics are only weakly dependent on
external shear \citep[e.g.][]{KKS97,Koo++09,Gav++12}.

At this stage we randomly draw important parameters such as the lens
magnitude, $g-i$ color, effective radius, ellipticity, and orientation
from a realistic distribution, and realize the lenses galaxies as
elliptically-symmetric de Vaucouleurs profile light distributions.
On top of them, fake lensed sources were
added. This approach has the advantage of being quite simple, and of
allowing direct identifications of factors limiting the completeness
of a lens survey. We anticipate that all the complexity of the lensing
galaxies, like their complex environment, the presence of a blue star
forming disk-like component or satellites, or gas rich minor mergers,
etc, would yield false positive signals in the \RF~pipeline. However,
with our simulations we can already set upper limits on the purity
that \RF~can achieve along with robust estimates of the survey
completeness.


\subsection{Observational aspects}

To be realistic, our simulations need to contain all the relevant
observational limitations that we face in real data. Although we are
focusing on the CFHTLS survey, our machinery is able to simulate a
wide variety of observational situations. Here, all the simulated
images are convolved with a CFHT/Megacam PSF that is constructed in
the same way as described in \S\ref{sect:method:residuals:PSF}. We
assume sky background surface brightnesses that are typical of CFHTLS
observing conditions, \ie~19.2 and 21.9 in the $i$ and $g$ bands
respectively. Exposure times are 5500 and 3500 s, respectively.
Although they are negligible here, readout noise and photon noise from
the lensed source are also included. Conversely, the photon noise from
the foreground deflector is carefully taken into account as we are
explicitly interested in the faint lensed features hidden beneath high
surface brightness foreground galaxies. 


\subsection{Statistics}\label{ssec:simstats}

We simulated \Nsim lines of sight, each exhibiting a deflector at the
center of coordinates and a source uniformly distributed within a circle of radius
$R_{\rm max}=3\arcsec$. Our statistics are boosted by avoiding	simulating
many foreground galaxies with no nearby background source. Therefore
numbers should be corrected by a factor  $\tau \equiv \pi R_{\rm
max}^2 / n_{\rm ng} \simeq 2 \times 10^{-4}$. Of course not all the
sources within this radius would give rise to substantial and hence
usable strong lensing, but some of these un-lensed sources could lead
to a positive \RF~detection signal and therefore should be included in
the simulation process. We avoid considering too large a value for
$R_{\rm max}$ because otherwise one could no longer neglect the
probability (as drawn from unclustered Poisson statistics) of 2 or
more sources being present within $R_{\rm max}$.

In addition, since we are dealing with extended sources, it is not
easy to build a criterion that would tell whether a lensing
configuration is giving rise to strong lensing or not.\footnote{One
could always imagine a very faint tail of surface brightness entering
the caustics of a lens even if most of the source's light is very far
away.} Therefore we chose to consider a certain level of total
magnification as the criterion for strong lensing being present. One
possible fiducial value is $\mu=2$, the total magnification reached
when a point-like source is on the edge of the multiple imaging region
of an SIS deflector. However, such a value is rather small, leading to
many occurrences of image systems that do not exhibit bright
counter-images and would therefore be of limited interest for lens
modeling. A more conservative value is $\mu=4$, which we find to imply
easily-identifiable multiple images. Unless otherwise stated, this is
the value we shall consider in what follows.


\subsubsection{Before applying \RF}

\begin{figure*}
 \includegraphics[width=0.99\linewidth]{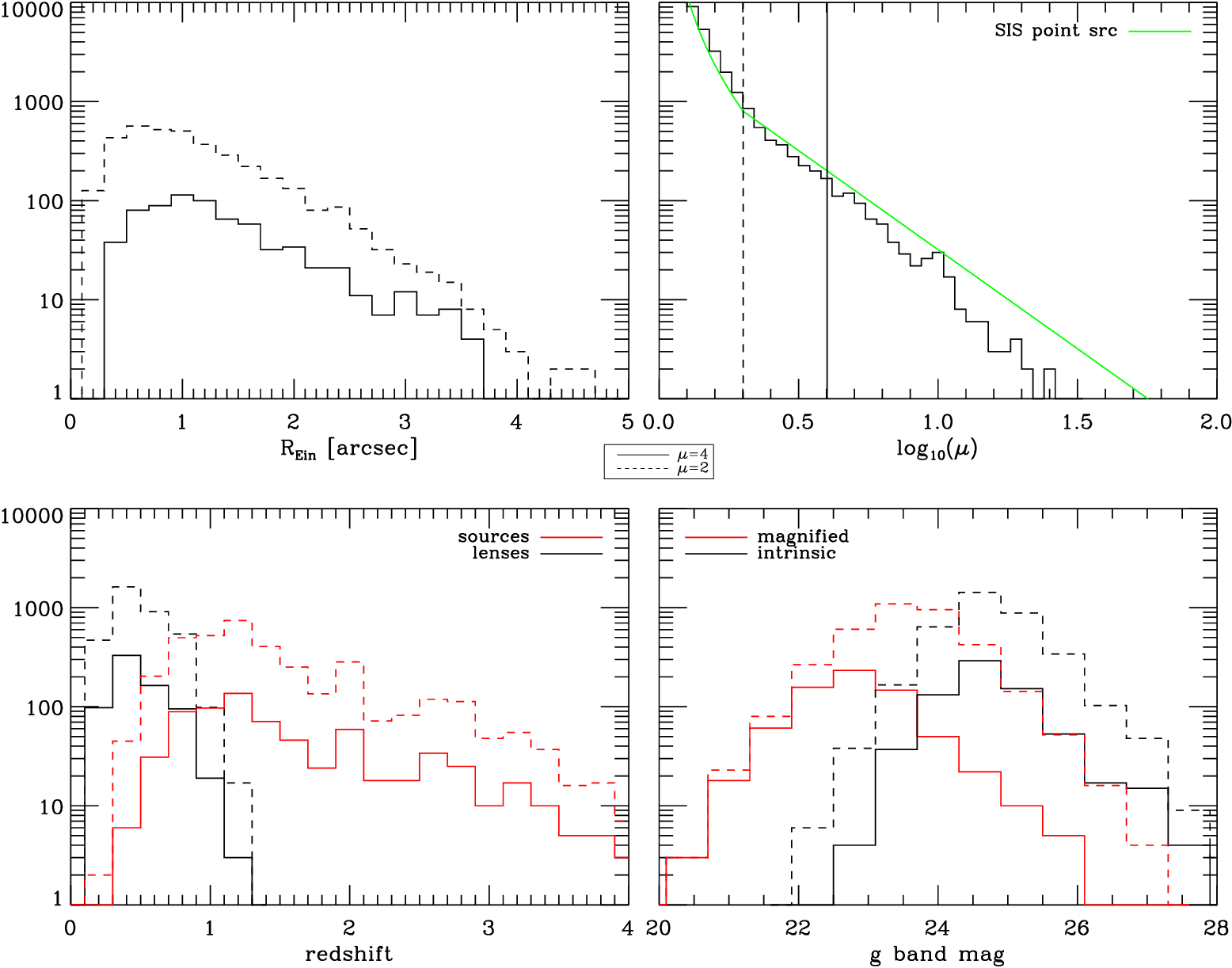}\\
 \caption{Distribution of various parameters for simulated lenses
satisfying a magnification $\mu\ge4$ (solid lines) and $\mu\ge2$ (dashed lines).
{\it Top left:\ \ } Einstein radius. {\it Top right:\ \ }
Magnification, with the two vertical lines marking the values $\mu=4$
and $\mu=2$. The green solid curve represents the predicted
magnification distribution for point sources lensed by a SIS mass
profile, for comparison. {\it Bottom left:\ \ } Redshift distribution
of the lenses (black) and sources (red). {\it Bottom right:\ \ }
Intrinsic (resp.\ magnified) $g$ band magnitude of the background
sources shown in black (resp.\ red). \label{fig:stat1}}
\end{figure*}

The first outcome of these simulations is shown in \Fref{fig:stat1}.
The distribution of Einstein radii is shown in the top left panel. The
median Einstein radius of the simulations is
$1\farcs17\mypm{0.87}{0.49}$. We stress that the statistics should not
be taken at face value for very large Einstein radii ($R_{\rm
Ein}\gtrsim 4\arcsec$), since the assumption of SIE mass distribution
should not apply to the few most massive galaxies at the centers of
groups or clusters of galaxies \citep[e.g.][]{New++13}. We also note
the rapid fall-off of the statistics at small $R_{\rm Ein}\lesssim
0\farcs3$ values. This is a clear consequence of the typical size of
sources having a half-light radius median value $0\farcs29$. This is
also seen in the top right panel where the differential probability
density of magnification does not scale as $\mu^{-3}$ at the high
magnification end, as it would for point-like sources, but the
probability instead decreases as the $\sim 3.5$th power of $\mu$
beyond $\mu \sim 10$.

For systems having $\mu\ge4$, the median deflector redshift is $z_{\rm
d}\simeq0.49\mypm{0.26}{0.14}$, while the median source redshift is
$1.28\mypm{1.2}{0.43}$ (bottom left panel of \Fref{fig:stat1}).
The magnified population of lensed arcs has a median value of $g\sim
22.8$ and corresponds to a median intrinsic magnitude of $g\sim
24.7$. 

In addition, the color index of the deflectors is $(g-i)_{\rm d}\simeq
2.0\pm0.4$ and the color index of the background sources lensed by
$\mu\ge4$ is $(g-i)_{\rm s}\simeq 0.76\mypm{0.79}{0.39}$. This
justifies our early hypothesis that background sources are much bluer
than the deflectors we consider here.

The top row of Table~\ref{tab:stat} lists the spatial density of
actual lenses that our simulations predict for the limiting source
magnitude $i<25$ and limiting deflector magnitude $i<22$. We expect
8.6 lenses per square degree magnified by $\mu=4$ or more.
These numbers compare well with the earlier
predictions of \citet{MBS05} or statistics from high-resolution
imaging data \citep{Fau++08,Jac08,Mar++09,NMT09,Paw++12}.

\begin{table}[!h]
\begin{center}
\caption{\label{tab:stat} Predicted global statistics of simulated
lenses.}
\begin{tabular}{l  ccc  ccc  }\hline\hline
   & \multicolumn{3}{c}{$\mu>4$} & \multicolumn{3}{c}{$\mu>2$} \\ \hline
  \# of existing lenses & \multicolumn{3}{c}{$8.6$} & \multicolumn{3}{c}{$44.3$} \\
  {\tt q\_flag}    &  $\ge0$  &  $\ge2$     & $\ge3$  &  $\ge0$  &  $\ge2$   & $\ge3$\\\hline
  \# of selected candidates & 12.5 & 6.4 & 2.5  & 12.5 & 6.4 & 2.5 \\
  \# of selected lenses   & 3.6 & 3.4 &  2.1 & 7.9 & 5.8 &  2.5\\
  completeness  (\%)& 42 & 39 & 25 & 18 & 13 & 6 \\
    purity   (\%)  & 29 & 53 & 84 & 63 & 91 & 100 \\\hline
\end{tabular}\end{center}
{\footnotesize Notes: Lensing events per square degree involving a
source brighter than $i=25$ and a foreground lensing ETG brighter than
$i=22$. The first row indicates the number of lenses predicted to
exist per square degree in the sky, the second row shows the number of systems
the \RF recovers, and the third row presents the number of actual
lenses among these. For each magnification threshold chosen as a
criterion for lensing, the ${\tt q\_flag}\ge0$ columns refers to the
statistics directly after the automated procedure while the  ${\tt
q\_flag}\ge2$ and  ${\tt q\_flag}\ge3$ refer to the quality level
assigned during visual classification. For each value of ${\tt q\_flag}$ we calculate completeness and purity at the ratio of first to third row and the ratio of the third to second row listing numbers per square degree.}
\end{table}

\begin{figure*}[!ht]
 \centering
 \includegraphics[width=0.99\linewidth]{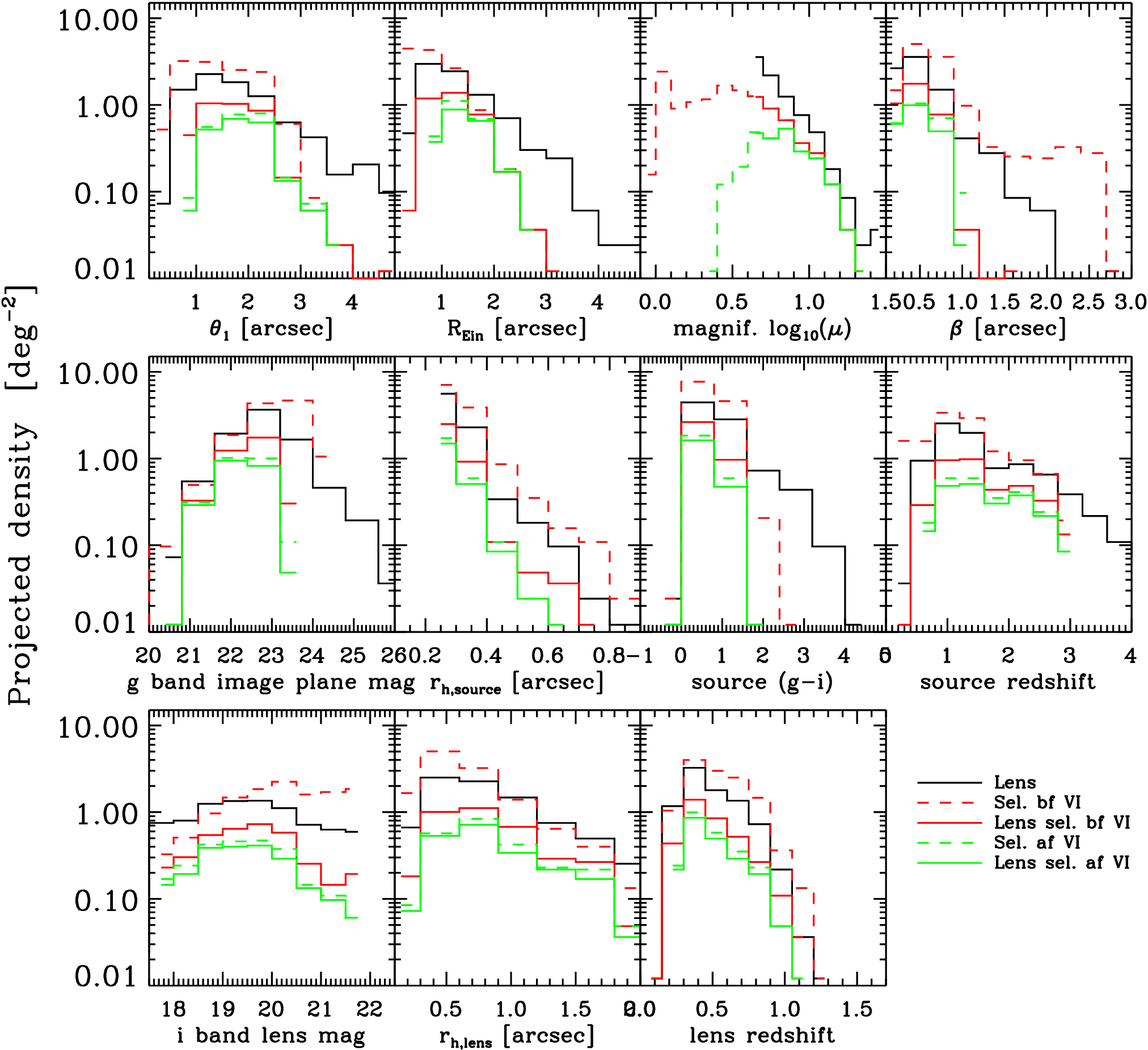}\\
 \caption{Statistics of recovery of simulated strong lenses for
 various sets of relevant parameters/observables within one square
 degree of the sky. The lenses are defined with the $\mu\ge4$
 criterion. From top left to bottom right, panels show the statistical
 dependency on the position of the furthest (and brightest) of the
 multiple images $\theta_1$,
 the Einstein Radius $R_{\rm Ein}$,
 the magnification $\mu$,
 the source position $\beta$,
 the arc (magnified) $g$ band magnitude,
 the source half-light radius,
 the source $(g-i)$ color index,
 the source redshift,
 the deflector $i$ band magnitude,
 the deflector half-light radius,
 and finally, the deflector redshift.
 In each panel, we show the distribution of all the lenses (solid
 black), the distribution of the candidates the algorithm
 automatically finds (dashed red), and among them, the distribution of
 the ones that are actually lenses. The additional loss of candidates
 produced by the subsequent level of selection (human inspection
 keeping only $\mathtt{q\_flag} \ge 3$ systems), is shown in green
 (solid for the actual lenses, and dashed for all the candidates).
 \label{fig:stat2}}
\end{figure*}


\subsubsection{After applying the automated part of \RF.}

The application of the \RF~pipeline with the settings presented in
\Sref{sect:method} will obviously change the above statistics. Not all
the lenses will be detected (loss of completeness) and some non-lenses
(in the sense $\mu\ge4$) will enter the sample (loss of purity). 
Before the visual inspection step detailed in
\Sref{sect:method:inspection}, \RF~yields the numbers shown in the
$\mathtt{q\_flag}\ge0$ columns of Table~\ref{tab:stat}\footnote{Although we mostly refer to results concerning the $\mu\ge4$ definition of a strong lensing event, we also report numbers related to the $\mu\ge2$ definition.}.
We can see that, per square degree, \RF~will
automatically detect 12.5 lens candidates. Among these, only 3.6
will be actual lenses magnified by $\mu\ge4$.
In other words, of the 8.6 lenses existing in a given square degree of the sky
(top row of \Tref{tab:stat}), 3.6 of them will be actual lenses,  these lenses being detected as
the same time as ($12.5-3.6=8.9)$ spurious non-lenses.
We thus conclude that the direct application of the automated
procedure will achieve a completeness of $3.6/8.6\simeq42\%$. Therefore, we see that the method
performs better for the most interesting lens systems. Conversely, we achieve a low purity
rate of $29\%$.

These global statistics can be better understood by viewing
\Fref{fig:stat2} where we overlay the distribution of some important
observable or hidden parameters for the population of lenses (solid
black), the population of recovered candidates (dashed red) and the
population of recovered true lenses (solid red). The ratio of the
solid red curve to the solid black curve should thus illustrate the
completeness, while the ratio of the solid red curve to the dashed
red curves gives the purity. In particular, we see that:
\begin{itemize}
\item The systems having their most distant lensed image lying at
radius $\theta_1$ in the range $1-2\farcs5$ are well recovered, and,
there, the purity is maximum. Beyond $\theta_1\sim 3 \arcsec$, the
\RF~radial exploration range would need to be changed in order to
catch these very few lenses. However we can extrapolate that many
false positives would also enter the detection sample, and would
therefore swamp the very few large separation lenses. On small
scales both purity and completeness are difficult to achieve for
$\theta_1< 1\arcsec$.
\item For a magnification $\mu>4$, the completeness does not change
much with $\mu$. Obviously, by construction, all the recovered
non-lenses are systems experiencing $\mu<4$.
\item Again, the source redshift has a very limited impact on the
recovery rate. We however notice that the low redshift $z_{\rm s}
\lesssim 1$ sources have a substantial contribution to the spurious
detections.
\item \RF~systematically misses the small red tail of the population of
sources; otherwise, the purity and completeness are quite constant in
the range $0\lesssim (g-i)_{\rm s}\lesssim1.2$.
\item The completeness and purity are maximized for the bright arcs
having $g<23$ and, at fainter magnitudes, many spurious system enter
the sample (but are not magnified much) and the completeness rapidly
falls off.
\item The source size only has a mild impact on purity and
completeness. We only see marginal evidence for the few large sources,
that cannot lead to high magnifications, contributing  to reducing the
purity of large arcs (that do result from high magnification).
\item The completeness is maximized for Einstein radii between $1$ and
$2\arcsec$. Below $1\arcsec$, the purity becomes poor, but it can
get close to unity for $R_{\rm Ein}\gtrsim 1\farcs6$. Conversely, the
completeness decreases for $R_{\rm Ein} \ge 2\arcsec$ and $R_{\rm
Ein}\lesssim 1\arcsec$ because of the limited analysis range of \RF,
as already noted for $\theta_1$.
\item Most of the spurious detections are due to sources with large
impact parameter $\beta>1\arcsec$ that lead to low magnifications.
Likewise, some actual lenses having a largely off-axis source,
presumably the ones with a large Einstein radius that we just saw we
are missing, are not recovered. These correspond again to the systems
with $\theta_1 \gtrsim 3\arcsec$, outside the \RF~exploration range.
Completeness and purity are both largest for the smallest impact
parameters $\beta \lesssim 0\farcs3$.
\item There is a mild selection effect with respect to deflector
redshift, given the parent population of simulated ETGs having $i<22$.
The deflector redshift has a very little impact on completeness and
purity slowly reduces for $\zd>0.4$.
\item \RF~does not imply particular selection effects with respect to
the apparent magnitude of the deflector, or their angular size due to
differential incompleteness. We however notice that the purity is
worse for the smallest and faintest deflectors because they are lower
mass or high redshift systems, and hence lead to lower magnifications.
\end{itemize}

\begin{figure}[!ht]
 \centering
\includegraphics[width=0.95\linewidth]{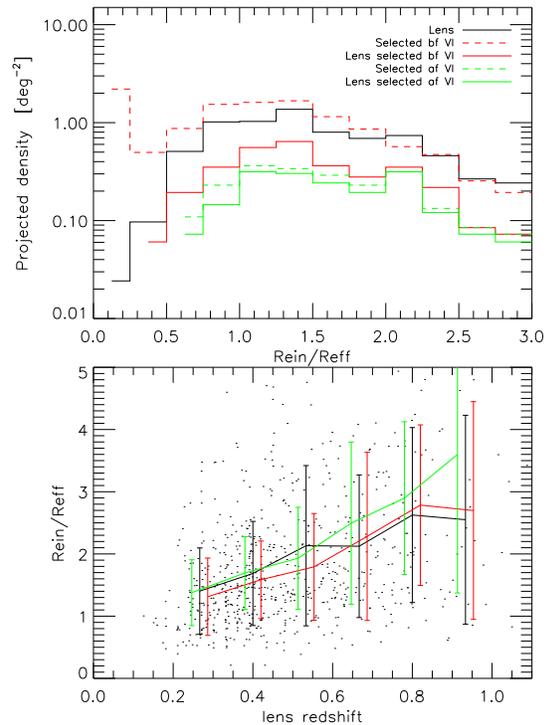}\\
 \caption{{\it Top panel: } Same as \Fref{fig:stat2} for the
distribution of $R_{\rm Ein}/R_{\rm eff}$. {\it Bottom panel:} Trend
with deflector redshift for the recovery of $R_{\rm Ein}/R_{\rm eff}$
with the same color coding as \Fref{fig:stat2}. Error bars
represent the $1-\sigma$ deviation about the mean. \label{fig:stat3}}
\end{figure}

In addition, it is important to check, for future scientific use of a
lens sample extracted from the \RF detection pipeline, that the
typical scale of detected Einstein radii is consistent with the parent
population of lenses at any redshift. \Fref{fig:stat3} shows that, in
our simulated lenses, there is no such particular selection effect as
a function of redshift. Studies of the redshift evolution of the deflectors'
properties will thus be more straightforward.

\subsubsection{After the subsequent visual classification}

The visual inspection step detailed in \Sref{sect:method:inspection}
will change the above statistics. If we consider only the best quality
flags systems as candidates, i.e., the ones with a
$\mathtt{q\_flag}\ge3$, the total purity can be increased
dramatically, to about $86\%$.This is  obviously at the expense of
completeness, which now reduces to $25\%$.  More statistics are
presented in the $\mathtt{q\_flag}\ge3$  and $\mathtt{q\_flag}\ge2$
columns of Table~\ref{tab:stat}. We can see that visually-classified
candidates with $\mathtt{q\_flag}\ge2$ already improve the purity to
$53\%$ while preserving the completeness $\sim 40\%$ of the automated
selection process. Therefore this first ``conservative'' visual
selection, which consists of keeping only systems with
$\mathtt{q\_flag}\ge2$ is of great value. Per square degree we expect
about $2.5$ (resp $6.4$) candidates with $\mathtt{q\_flag}\ge3$ (resp.
$\mathtt{q\_flag}\ge2$), which corresponds to a $80\%$ (resp. $50\%$)
decrease as compared to before visual classification.

As a function of relevant parameters, the change in statistics
for$\mathtt{q\_flag}\ge3$ is shown in the panels of \Fref{fig:stat2}
as green histograms, the solid one showing the recovered lenses with
$\mu\ge4$ and the dashed one showing all the recovered candidates. We
see that the inspection is particularly efficient at removing the
low-$R_{\rm Ein}$ spurious systems, while preserving the high-$R_{\rm
Ein}$ lenses. Likewise, the many spurious candidates with low
magnification and large impact parameter are correctly discarded at
low extra completeness cost. This is also true for the spurious faint
arcs, the low redshift sources, and spurious small and faint
deflectors.

Regarding selection effects in terms of the typically probed physical
scale $R_{\rm Ein}/R_{\rm eff}$, we see in \Fref{fig:stat3} that no
significant change is introduced with respect to the parent population
of lenses.


\subsubsection{Selection effects related to the photometric preselection.}
\label{sssec:selcol}

The above steps of the simulations are based on a pure and complete
(down to $i<22$) parent population of ideal ETGs that can subsequently
be tested for the presence of strongly lensed arc-like features. This
is an idealistic case that we now call into question. The presence of
another object along the line of sight of an ETG, whatever its
redshift, may perturb the photometry of the ETG. This was anticipated
in the presentation of the \RF pipeline in
\Sref{sect:method:selection}.

The presence of the secondary object, either magnified or not, will
perturb the photometry in several ways. At the catalog level
(produced by {\tt SExtractor} for the CFHTLS), if the secondary object
is bright enough and close enough, the source extractor will be fooled
by the secondary in one or several bands. This can lead to a
misidentification if the secondary is of similar flux or brighter than
the deflector. Our simulations suggest that $\sim3\%$ of all the
simulated sightlines lead to such misidentifications. Furthermore the
frequency rises to $\sim 6\%$ of the sightlines that involve a
$\mu\ge4$ magnification event. Therefore we readily see that 6\% of
the strong lenses we simulated could be lost at the ETGs catalog
level.

\begin{figure}[!ht]
 \centering
 \includegraphics[width=0.99\linewidth]{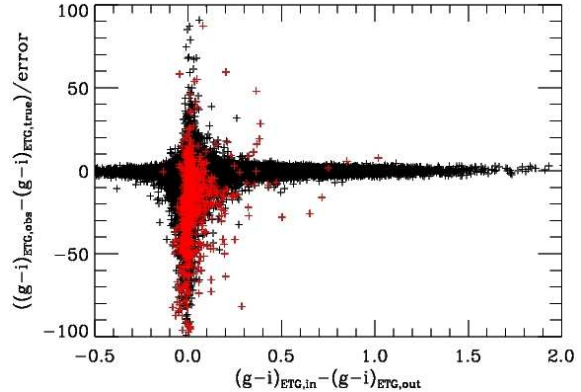}\\
 \caption{Difference in $(g-i)$ color index for simulated lines of
 sight involving an ETG and another random object (black crosses).
 Actual lenses yielding $\mu \ge 4$ are shown in red. The difference is
 expressed in units of the photometric error and as a function of the
 color gradient measured in two $R_{\rm in}=1\farcs86$ and $R_{\rm
 out}=3\farcs35$ apertures as in \Sref{sect:method:selection}.
 \label{fig:colsel}}
\end{figure}

The photometry will not be affected as dramatically, though
photometric redshifts can be altered, since even a small amount of
flux coming from the lensed object will modify the SED of the blended
\{foreground ETG, lensed arc\} system, possibly leading to a bias in the
redshift estimate if taken as a unique object in the parent catalog.
This can lead to a loss of lensing ETGs, since photometric redshifts
are used in our preselection.
Figure~\ref{fig:colsel} shows the difference of measured and input
$(g-i)$ color indices divided by the photometric error on it, as a
function of the color gradient defined in \Eref{eq:colsel2}. We
can see strong departures from the input color index that will presumably
lead to perturbations in SED fitting, hence implying unreliable
photometric redshifts or spectrophotometric template types.
We see that the lenses with $\mu \ge 4$ clearly exhibit different colors.
More quantitatively, we estimate that about 5\% of our simulated lines
of sight will end up in photometric catalogs in which the $(g-i)$
color index will depart by more than $0.2$ magnitudes from the
intrinsic color of the foreground galaxy, leading to misleading SED
fits. The fraction of spurious colors increases to $\sim 38\%$ for the
populations of $\mu \ge 4$ actual lenses. This has therefore to be
accounted for when dealing with a photometric redshift catalog. This
was the motivation of our conservative color gradient cuts of
Equations~\ref{eq:colsel1} and \ref{eq:colsel2}. By ignoring SED
fits to the objects satisfying these criteria, we are able to mitigate
the problem and limit the loss of $\mu\ge4$ lenses to about 19\% with
little dependency on deflector or source parameters, except a mild
bias against the smallest Einstein radii leading to apparently blue
core foreground galaxies. This is an illustration of the limitations
of lens-oriented surveys that have difficulties disentangling the
light from the foreground and the background objects when they emit at
similar wavelengths.

\renewcommand{\arraystretch}{1.10} 
\begin{deluxetable}{lrrcccccc}
\tablewidth{0pt}
\tabletypesize{\small}
\tablecaption{Additional SL2S lens candidates.}
\tabletypesize{\footnotesize}
\tablehead{
\colhead{name SL2SJ...} &
\colhead{RA} &
\colhead{DEC} &
\colhead{mag$_i$} &
\colhead{$z_{\rm phot,d}$} & 
\colhead{$z_{\rm d}$} &
\colhead{$z_{\rm s}$} &
\colhead{\tt q\_flag} &
\colhead{\tt confirmed}
}
\startdata
020833$-$071414 &    32.1379 &    -7.2372 & 18.11 & 0.424 & 0.428 & \nodata & 3 & 3 \\
021325$-$074355 &    33.3522 &    -7.7319 & 19.56 & \nodata & 0.717 & 3.480 & 3 & 3 \\
022511$-$045433 &    36.2960 &    -4.9093 & 16.92 & 0.253 & 0.238 & 1.200 & 3 & 3 \\
022610$-$042011 &    36.5444 &    -4.3366 & 18.85 & 0.542 & 0.494 & 1.230 & 3 & 3 \\
022648$-$040610 &    36.7016 &    -4.1029 & 20.16 & 0.815 & 0.766 & \nodata & 3 & 3 \\
022940$-$040639 &    37.4183 &    -4.1109 & 17.98 & 0.290 & \nodata & \nodata & 3 & \nodata \\
023251$-$040823 &    38.2149 &    -4.1399 & 19.02 & 0.510 & 0.352 & 2.340 & 3 & 3 \\
090407$-$005952 &   136.0330 &    -0.9980 & 20.15 & 0.724 & 0.611 & 2.360 & 3 & 3 \\
095921+020638 &   149.8407 &     2.1107 & 20.28 & 0.455 & 0.552 & 3.350 & 3 & 3 \\
135847+545913 &   209.6958 &    54.9870 & 19.38 & 0.551 & 0.510 & \nodata & 3 & 3 \\
142003+523137 &   215.0142 &    52.5272 & 20.99 & 0.510 & 0.390 & 1.410 & 3 & 3 \\
221045$-$005918 &   332.6892 &    -0.9884 & 21.62 & 1.024 & \nodata & \nodata & 3 & 2 \\
221326$-$000946 &   333.3591 &    -0.1629 & 19.71 & 0.307 & 0.338 & 3.450 & 3 & 3 \\
020850$-$034459 &    32.2110 &    -3.7500 & 18.84 & 0.455 & \nodata & \nodata & 2 & \nodata \\
020905$-$090155 &    32.2715 &    -9.0322 & 19.80 & 0.604 & \nodata & \nodata & 2 & \nodata \\
021004$-$063011 &    32.5205 &    -6.5032 & 19.36 & 0.663 & \nodata & \nodata & 2 & \nodata \\
021234$-$083325 &    33.1442 &    -8.5571 & 20.38 & 0.715 & \nodata & \nodata & 2 & \nodata \\
021300$-$084310 &    33.2527 &    -8.7196 & 19.26 & 0.471 & \nodata & \nodata & 2 & 0 \\
021604$-$045855 &    34.0177 &    -4.9820 & 18.68 & 0.574 & \nodata & \nodata & 2 & \nodata \\
021619$-$062958 &    34.0820 &    -6.4996 & 18.84 & 0.378 & \nodata & \nodata & 2 & \nodata \\
021620$-$044003 &    34.0872 &    -4.6676 & 20.23 & 0.683 & \nodata & \nodata & 2 & 1 \\
022056$-$074311 &    35.2352 &    -7.7199 & 20.48 & 0.688 & \nodata & \nodata & 2 & \nodata \\
022115$-$090602 &    35.3128 &    -9.1006 & 19.21 & 0.475 & \nodata & \nodata & 2 & \nodata \\
022203$-$054429 &    35.5149 &    -5.7416 & 19.69 & 0.753 & \nodata & \nodata & 2 & \nodata \\
022233$-$083728 &    35.6380 &    -8.6246 & 18.86 & 0.663 & \nodata & \nodata & 2 & \nodata \\
022241$-$053851 &    35.6734 &    -5.6477 & 19.50 & 0.502 & \nodata & \nodata & 2 & 1 \\
022508$-$041749 &    36.2847 &    -4.2970 & 21.13 & 0.719 & 0.747 & \nodata & 2 & 1 \\
022612$-$044055 &    36.5518 &    -4.6822 & 21.89 & 1.072 & 0.977 & \nodata & 2 & 0 \\
022621$-$040054 &    36.5891 &    -4.0152 & 20.08 & 0.672 & \nodata & \nodata & 2 & \nodata \\
022648$-$090421 &    36.7023 &    -9.0726 & 18.29 & 0.492 & 0.456 & \nodata & 2 & 3 \\
022733$-$075021 &    36.8879 &    -7.8394 & 20.57 & 0.420 & \nodata & \nodata & 2 & \nodata \\
022757$-$042203 &    36.9864 &    -4.3681 & 20.10 & 0.788 & \nodata & \nodata & 2 & \nodata \\
022929$-$040013 &    37.3715 &    -4.0037 & 19.87 & 0.641 & \nodata & \nodata & 2 & \nodata \\
023051$-$082422 &    37.7140 &    -8.4064 & 19.66 & 0.586 & \nodata & \nodata & 2 & \nodata \\
023251$-$044827 &    38.2150 &    -4.8076 & 20.61 & 0.320 & \nodata & \nodata & 2 & \nodata \\
023252$-$043026 &    38.2180 &    -4.5073 & 18.86 & 0.450 & \nodata & \nodata & 2 & \nodata \\
023402$-$051950 &    38.5098 &    -5.3307 & 19.59 & 0.466 & 0.450 & 0.330 & 2 & 0 \\
023413$-$055715 &    38.5563 &    -5.9543 & 19.09 & 0.484 & \nodata & \nodata & 2 & \nodata \\
023506$-$082049 &    38.7760 &    -8.3472 & 19.13 & 0.548 & \nodata & \nodata & 2 & \nodata \\
084612$-$023751 &   131.5535 &    -2.6311 & 19.12 & \nodata & \nodata & \nodata & 2 & 0 \\
084909$-$041226 &   132.2896 &    -4.2074 & 20.36 & 0.797 & 0.722 & 1.540 & 2 & 3 \\
085731$-$010404 &   134.3794 &    -1.0679 & 18.72 & 0.660 & \nodata & \nodata & 2 & \nodata \\
085938$-$010213 &   134.9088 &    -1.0370 & 19.22 & 0.604 & 0.520 & 0.510 & 2 & 0 \\
090106$-$025906 &   135.2752 &    -2.9852 & 21.19 & 0.844 & 0.670 & 1.190 & 2 & 2 \\
090116$-$020541 &   135.3197 &    -2.0948 & 20.99 & 0.678 & \nodata & \nodata & 2 & 1 \\
090119$-$021039 &   135.3332 &    -2.1777 & 19.51 & 0.428 & \nodata & \nodata & 2 & 1 \\
090327$-$015905 &   135.8627 &    -1.9849 & 20.11 & \nodata & \nodata & \nodata & 2 & 0 \\
095850+023052 &   149.7091 &     2.5147 & 19.15 & 0.472 & \nodata & \nodata & 2 & 0 \\
100108+024029 &   150.2850 &     2.6749 & 19.03 & 0.378 & \nodata & \nodata & 2 & 1 \\
100141+020444 &   150.4217 &     2.0789 & 19.34 & 0.783 & \nodata & \nodata & 2 & 0 \\
100220+022335 &   150.5845 &     2.3933 & 20.74 & 0.600 & \nodata & \nodata & 2 & 1 \\
135629+545423 &   209.1235 &    54.9064 & 19.38 & 0.550 & \nodata & \nodata & 2 & \nodata \\
135702+561000 &   209.2589 &    56.1669 & 19.95 & 0.606 & \nodata & \nodata & 2 & \nodata \\
135915+554853 &   209.8150 &    55.8149 & 19.51 & 0.403 & \nodata & \nodata & 2 & 0 \\
140315+532540 &   210.8148 &    53.4279 & 20.20 & \nodata & \nodata & \nodata & 2 & 0 \\
140414+514414 &   211.0588 &    51.7374 & 19.69 & \nodata & \nodata & \nodata & 2 & \nodata \\
140425+520506 &   211.1062 &    52.0850 & 18.82 & 0.522 & \nodata & \nodata & 2 & \nodata \\
140606+553047 &   211.5272 &    55.5131 & 21.11 & 0.620 & \nodata & \nodata & 2 & 0 \\
140650+522619 &   211.7097 &    52.4386 & 20.05 & 0.859 & 0.716 & 1.470 & 2 & 3 \\
141143+535043 &   212.9318 &    53.8454 & 18.18 & 0.434 & 0.393 & \nodata & 2 & \nodata \\
141336+531235 &   213.4025 &    53.2099 & 20.12 & 0.407 & \nodata & \nodata & 2 & \nodata \\
141917+511729 &   214.8219 &    51.2913 & 18.72 & 0.468 & \nodata & \nodata & 2 & 3 \\
142115+525137 &   215.3130 &    52.8605 & 18.25 & 0.407 & \nodata & \nodata & 2 & \nodata \\
142147+563052 &   215.4476 &    56.5147 & 18.43 & 0.333 & \nodata & \nodata & 2 & \nodata \\
142321+572243 &   215.8413 &    57.3786 & 19.11 & 0.611 & \nodata & \nodata & 2 & 2 \\
142935+530819 &   217.3973 &    53.1386 & 20.01 & 0.555 & \nodata & \nodata & 2 & \nodata \\
143002+554634 &   217.5081 &    55.7763 & 19.84 & 0.645 & \nodata & \nodata & 2 & 1 \\
143255+551447 &   218.2329 &    55.2466 & 18.54 & 0.400 & 0.423 & \nodata & 2 & \nodata \\
221606$-$175131 &   334.0286 &   -17.8588 & 20.51 & 0.937 & 0.860 & \nodata & 2 & 2 \\
221610+002116 &   334.0419 &     0.3546 & 19.68 & 0.733 & 0.751 & \nodata & 2 & \nodata \\
222055+011825 &   335.2296 &     1.3070 & 18.31 & 0.305 & \nodata & \nodata & 2 & \nodata \\

\enddata
\tablecomments{\label{tab:cands2} This additional table lists candidates that were
found serendipitously with earlier implementations of \RF~or in the
CFHTLS Deep during the development stages. Columns share the definition of table~\ref{tab:cands}.}
\end{deluxetable}


\section{Follow-up observations}\label{sect:followup}  

In order to assess the merits of the \RF~procedure, we undertook a
series of follow-up observations of our sample of candidate strong
lensing events. The missing pieces of evidence for validating a
candidate as a strong lens are the knowledge of the source and
deflector redshifts along with some high spatial resolution imaging
that would unambiguously show signatures of strong lensing, for example with the
multiplicity of sources of similar morphology and color.


\subsection{High-resolution imaging with \hst}\label{sect:followup:HST}

The most important follow-up effort we carried out was the imaging of
our lens candidates over the course of three \hst~cycles as part of
the SNAPSHOT programs 10876, 11289 (PI Kneib) and 11588 (PI Gavazzi).
The technical details of the observations and their reduction is
presented in \citet{Gav++12} and \citet{PaperIII}. The earliest
observations consisted in F814W snd F606W exposures summing to a
single-orbit visit. Subsequently, the failure of ACS led us to conduct
single filter (F606W) 1200-second observations of some systems with
WFPC2, while the most recent observations were performed with the WFC3
camera using the F600LP and F475X wide filters to fill a full
\hst~orbit.

Besides these dedicated observations, we also took advantage of the
existence of the COSMOS \citep{Sco++07} survey and the Extended Groth
Strip (EGS) data that happen to coincide with the CFHTLS D3 and D1
fields respectively. The known lenses in these \hst~data
\citep{Mou++07,Fau++08,Jac08} were used at various stages of the
developement of \RF~to help tune the free parameters of the pipeline
such that the number of recovered COSMOS and EGS lenses would be
maximized. The lenses used at that time were not subsequently
considered as being part of the statistically homogeneous list of
candidates presented in Table~\ref{tab:cands}, and were extracted from
the CFHTS Wide fields' data only.

Of the 19 main sample candidates of Table~\ref{tab:cands} that have
been observed with \hst, 16 turned out to be definite lenses ({\tt
confirmed=3}), 1 a probable lens({\tt confirmed=2}), 2 possible
lenses({\tt confirmed=1}) and 1 non lens ({\tt confirmed=0}). The
confirmation rate for the final implementation of \RF~is high.

During the development phase of \RF, we intentionally dedicated our
follow-up to a broad range of quality flag value targets (from {\tt
q\_flag}=0 to {\tt q\_flag}=3).  Of the first 43 candidates observed
with \hst during this development phase,  14 systems were subsequently
ranked with {\tt confirmed=3}, 2 with {\tt confirmed=2}, 10 with {\tt
confirmed=1} and 17 with {\tt confirmed=0}. Defining good quality
additional candidates to have  {\tt q\_flag}$\ge 2$, the confirmation
rate provided by \hst~in this phase was 50\%. This can be compared
with the confirmation rate of about $89\%$ that we eventually achieved
in the final main sample, having learned from the first series of
\hst~observations. 


\subsection{High-resolution imaging with Keck Laser Guide Star
Adaptive Optics}\label{sect:followup:LGSAO}

The Keck Adaptive Optics system has been proven to be very effective
for the study of strong lensing systems, routinely delivering a stable
enough PSF to enable a variety of science applications ranging from
lens confirmation \citep{Tre++11} to lens modeling
\citep{Mar++07,Gav++11,Aug++11,Bre++12,Fu++12,Lag++12,Veg++12} especially when the sources are red or
the deflectors are dusty.

In the early stages of the follow-up effort, we obtained high
resolution imaging ground based K-band images using the NIRC2 Camera
on the Keck II Telescope assisted by Laser Guide Star Adaptive Optics.
Observations were conducted in September 2007 under good observing
conditions, sub-arcsecond natural seeing. The AO system worked well
delivering typical Strehl ratio of approximately 0.2.  Given the
early-stage of \RF, the aim of this campaign was limited to
confirming/rejecting candidate lenses using quick ``snapshot''
exposures of approximately 30 minutes integration, and help in the
refinement of the classification criteria.

In total 9 SL2S targets were observed, including 2 that later ended up
in the main SL2S sample. One of the two main sample candidates ended
up being a definite lens. The second one remains inconclusive ({\tt
confirmed=1)} even with additional WFPC2 observations. Of the 7
non-main sample candidate 5 were identified as contaminants using the
AO images. No conclusion could be drawn on the remaining two (one {\tt
confirmed =1} and one {\tt confirmed=2}). The contaminants were
revealed to be spiral galaxies by the AO images.  Thus, although no
system was confirmed using the AO data alone, the Keck data helped
identify spiral contaminants and refine our selection process to
increase success rate in subsequent iterations of \RF.


\subsection{Spectroscopic follow-up}\label{sect:followup:spec}
The SL2S spectroscopic campaign was started in 2006. 
The goal of our spectroscopic observations was to measure the deflector and source redshifts and deflector velocity dispersion for all our systems. The latter quantity is not required to assert that a lens candidate really is a deflector but it was used as an additional constraint on the mass properties of the confirmed lenses \citep{Ruf++11,PaperIV}.
Different telescopes (Keck, VLT and Gemini), instruments (LRIS, DEIMOS, X-Shooter\footnote{ESO/VLT programs 086.B-0407(A) and 089.B-0057(A), PI Gavazzi}, GNIRS) and setups have been used to achieve this goal, reflecting technical advances during the years and the optimization of our strategy.
The procedure we used to analyze the spectra and measure redshifts along with velocity dispersions is presented in great detail by \citet{PaperIV}. Since this publication, we obtained 4 redshift measurements from a new XSHOOTER program\footnote{ESO/VLT program 092.B-0663(A), PI Gavazzi} in Fall 2013. They are included in the summary tables \ref{tab:cands} and \ref{tab:cands2}. A detailed description of these data is left for a future work (Sonnenfeld et al., in prep).

The most difficult quantity to measure was the redshift of the lensed source which often required near-IR coverage to detect the important redshifted emission lines.

In addition, 31 lens candidates were bright enough to have an entry in the SDSS3 public catalog \citep[9th data release,][]{ahn++12}. 18 belong to the main sample. None of these fiber spectra could yield a measurement of the source redshift because the signal-to-noise ratio was too low.

Among the systems that we spectroscopically followed up, 40 were part of the main sample of table~\ref{tab:cands}, of which 31 (resp. 9) are definite {\tt confirmed=3} (resp. probable {\tt confirmed=2}) lenses. Another 16 spectra were obtained for the additional candidate list. 15 of them are definite lenses.

Altogether we followed up 56 lens candidates. All but one allowed us to measure the deflector redshift. 51 observations yielded a measurement of the source redshift.
For 6 of these 51 redshift estimates, the corresponding redshift coincides with the redshift of the foreground deflector. They were thus flagged with {\tt confirmed=0}.


\subsection{Confirmation rate and merits of the classification scheme}\label{sect:classif}

We can now compare the overall quality of our selection and visual
classification scheme with the new classification allowed by the
availability of redshifts or high-resolution imaging. In
table~\ref{tab:compar1}, we show the number of candidates that were
classified with a given {\tt q\_flag} score and, subsequently,
after additional spectroscopic or imaging data, were
reclassified with a given {\tt confirmed} value according to our
already defined scheme: 3 definitely a lens, 2 probably a lens, 1
possibly a lens, and, 0 not a lens.

We see that the correspondence between {\tt q\_flag} and {\tt confirmed} is very tight
at the extremes, either a value 3 for "definitely a lens" or a value 0 for "definitely not a lens".
In between, the classification is more ambiguous. The {\tt q\_flag}=2 systems end up with about the same frequency into all
possible values of {\tt confirmed}. With time, our visual classification skills increased.
Most of the systems having {\tt q\_flag}=2 and {\tt confirmed}$<2$ come from early \RF implementations.
Indeed, only 2 out of 21 such systems belong to the main lens sample of Table~\ref{tab:cands}.

The confirmation rate ({\tt confirmed}=3) for {\tt q\_flag}$>2$ in the main sample is about 66\%.
Extrapolating to the total number of lens candidates in the main sample,
we conclude that our sample should contain about 220 lenses.

\begin{table}
\begin{center}
 \caption{Comparison of pre- and post- follow-up observation ranking. 
\label{tab:compar1}}
 \begin{tabular}{ c | c c c c}\hline\hline
{\tt q\_flag} & \multicolumn{4}{c}{ {\tt confirmed} } \\ 
{}   &  0  & 1 & 2 & 3\\ \hline
0 & $  6 $ & $  1 $ & $  0 $ & $  0 $ \\ 
1 & $  7 $ & $  4 $ & $  0 $ & $  1 $ \\ 
2 & $ 11 $ & $ 10 $ & $  7 $ & $ 11 $ \\ 
3 & $  2 $ & $  0 $ & $  5 $ & $ 37 $ \\\hline
\end{tabular}

\end{center}
{\footnotesize Notes: number of followed-up candidates with a given
{\tt q\_flag} and {\tt confirmed} flags.}
\end{table}

\subsection{Statistical comparison with simulations}\label{sect:results}

In this section, we investigate whether the conclusions that can be drawn from the follow-up observations
of our lens candidates are consistent with the predicted statistics suggested by our simulations of
\S~\ref{sect:simus}.

First, accounting for the $\sim20\%$ loss of completeness expected from photometric perturbations of 
deflectors detailed in \S~\ref{sssec:selcol}, we should update the statistics of strong lensing events
predicted in \S~\ref{ssec:simstats}. By doing so, we estimate that, over the \SArea deg$^2$ of the CFHTLS imaging
data, about 1040 lenses should be present down to the simulated depth. \RF should detect about 960 systems
with {\tt q\_flag}$\ge2$, of which $\sim$ 410 should be actual $\mu\ge4$ lenses. This somewhat over-predicts
by a factor of $\sim2$ the actual number of {\tt q\_flag}$\ge2$ candidates that we found.

The predictions for simulations with {\tt q\_flag}$\ge3$ are in much better agreement
with our actual observed numbers. Those differences suggest that our classification scheme
might have been slightly more strict in the observations. This illustrates the limitations of a visual classification that hampers any very accurate calibration. Keeping also in mind that our simulations do not
incorporate all the complexity of a fully realistic surface brightness profile that could exhibit some blue star
forming regions in the deflector plane, satellites and, more generally, small amounts of color gradients,
we conclude that getting numbers within a factor of $\lesssim 2$, depending on the stability of a
{\tt q\_flag=2} or {\tt q\_flag=3} ranking, is already quite satisfactory.


\section{Summary and conclusions}\label{sect:concl}

We have developed and implemented a novel method for the automated
detection and classification of strong galaxy-scale gravitational
lenses in ground-based imaging surveys. The method is based on
difference imaging between blue and red filters, automated
morphological cuts, and final visual inspection. The alogorithm has
been extensively tested in fields containing known lenses and via
simulated strong lensing configuration to characterize its selection
function, purity and completeness. The algorithm has been applied to
search for strong lens candidates in the $\sim \SArea$ deg$^2$ of CFHTLS
imaging. The results of the search as well as of extensive follow-up
for confirmation and rejection of candidates have been presented. Our
main results can be summarized as follows.

\begin{itemize}

\item The \RF algorithm is highly efficient, 
retaining just \NRF lens candidates (\XRF) from a sample of \NETGs
pre-selected early-type galaxies. This corresponds to about 18
candidates per square degree.

\item Hierarchical visual inspection of these objects led to the definition of a sample of \Ncands candidates of sufficient quality $({\tt
q\_flag}\ge2)$ for follow-up observation.  These systems are listed in
table~\ref{tab:cands}.  This procedure took approximately 48
person-hours to complete, or $\lesssim30$ person-minutes per square
degree. The follow-up campaigns allowed us to confirm 33 definite gravitational lenses in this (main) sample.
Extrapolating to the full catalog, about 220 lenses are still to be confirmed in our sample.

\item During the early stages of this work, while still tuning the parameters of the \RF~algorithm, we also gathered \NfX~lens candidates with {\tt q\_flag}$\ge2$ that did not remain in the statistical main sample of \Ncands~systems. 16 of those systems have been confirmed as lenses. The full list of additional candidates is given in table~\ref{tab:cands2}.

\item These SL2S galaxy-scale lenses lie in the redshift range 
$0.3\lesssim z \lesssim 0.8$, and thus complement well the SLACS
galaxy-scale lens sample at median redshift 0.2. 

\item High-resolution imaging and extensive spectroscopic follow-up 
candidates allowed \Nlens (+\Nlensn systems from the additional non
statistical sample) new gravitational lenses to be confirmed. The
analysis of the confirmed sample has been carried out in
\citep{Gav++12}, \citep{Ruf++11}, \citep{PaperIII} and \citep{PaperIV} and yielded
valuable constraints on the evolution of the mass profile of massive
ETGs back to redshift $z\sim0.8$ thanks to a combined lensing+dynamics
analysis.

\item We tested the response of \RF to simulated images of ETGs exhibiting a range of detectable
features along the line of sight to quantify the fraction of recovered true lenses as well as the
contamination of our lens sample by non-lenses. Both statistics are in broad agreement
(within a factor $<2$) given the limitations of the simulations and the imprecision of the visual
classification at the end of the \RF procedure. A rough number of 1.5 definite lenses per sq. degree
can be obtained with a method similar to \RF on data similar to the CFHTLS.

\item Those simulations allowed us to check that the overall completeness we could achieve for such a survey after visual inspection is close to 40\%. However, restricting the search to a narrow range of Einstein radii, between 0.9 and 2 arcsec, as well as arcs (or lensed features) brighter than a magnitude of 22.5 in the $g$ band, can significanly increase the completeness to a value $\sim$70\%. Lenses in a different regime could be better seeked with other surveys (deeper and better resolution for small Einstein radii and faint sources or wider for large Einstein radii).

\end{itemize}

This paper represents a step forward in the detection of galaxy-scale
gravitational lenses in wide field ground-based imaging surveys. 
Upcoming (such as DES, KiDS and HSC) and future imaging datasets (such
as Euclid and LSST) will cover thousands of square degrees of sky and
are expected to contain a few hundred thousand such lenses
\citep{LSST,Euclid}. Our ability to detect them efficiently in large
numbers will depend largely on the development of algorithms such as
\RF. However, the current version of \RF would require too much human
supervision ($\sim$5000 person hours scaling from CFHTLS) for an
individual investigator. Therefore a successful lens search in such
surveys will require either the development of additional automated
stages of classification to increase the purity of lens candidates by
at least an order of magnitude, or the implementation of
crowd-sourcing to carry out the visual inspection. Both approaches are
currently being pursued by the authors: one strategy for the former is
to perform a fast lens modeling analysis as the final stage of
selection
\citep{Mar++09}, while the latter approach is currently being
investigated by the Space Warps project (Marshall et al, in
preparation) using the same CFHTLS dataset as described here.


\acknowledgments
We thank the members of the SL2S collaboration for many useful
discussions.
This work is based on observations obtained with MegaPrime/MegaCam, a
joint project of CFHT and CEA/DAPNIA, at the Canada-France-Hawaii
Telescope (CFHT) which is operated by the National Research Council
(NRC) of Canada, the Institut National des Sciences de l'Univers of
the Centre National de la Recherche Scientifique (CNRS) of France, and
the University of Hawaii. This work is based in part on data products
produced at TERAPIX and the Canadian Astronomy Data Centre as part of
the Canada-France-Hawaii Telescope Legacy Survey, a collaborative
project of NRC and CNRS.
This research is also based on XSHOOTER observations made with ESO Telescopes at the Paranal
Observatory under programme IDs 086.B-0407(A), 089.B-0057(A) and 092.B-0663(A).
RG acknowledges support from the Centre National des Etudes Spatiales
(CNES) and the Programme National Cosmologie et Galaxies (PNCG).
PJM acknowledges support from the Royal Society in the form of a university
research fellowship.
TT acknowledges support from the NSF through CAREER award NSF-0642621, and from
the Packard Foundation through a Packard Research Fellowship.
AS acknowledges support by a UCSB Dean Graduate Fellowship.
This research is supported by NASA through Hubble Space Telescope
programs GO-10876, GO-11289, GO-11588 and in part by the National
Science Foundation under Grant No. PHY99-07949, and is based on
observations made with the NASA/ESA Hubble Space Telescope and
obtained at the Space Telescope Science Institute, which is operated
by the Association of Universities for Research in Astronomy, Inc.,
under NASA contract NAS 5-26555, and at the W.M. Keck Observatory,
which is operated as a scientific partnership among the California
Institute of Technology, the University of California and the National
Aeronautics and Space Administration. The Observatory was made
possible by the generous financial support of the W.M. Keck
Foundation. 
%


\nocite{RGO99,Fas++04}


\end{document}